\documentclass[preprint]{emulateapj}
\usepackage[flushleft]{threeparttable}
\usepackage{color}
\usepackage{longtable}
\usepackage{amsmath,natbib,graphicx}
\usepackage{graphics,graphicx}
\usepackage{epsf}
\usepackage{epstopdf}
\input{epsf}
\usepackage{epsfig}
\usepackage{color}
\DeclareGraphicsExtensions{.jpg,.pdf,.png,.eps,.ps}
\usepackage{hyperref}
\usepackage{ulem}
\hypersetup{
    bookmarks=true,         
    unicode=false,          
    pdftoolbar=true,        
    pdfmenubar=true,        
    pdffitwindow=false,     
    pdfstartview={FitH},    
    pdftitle={My title},    
    pdfauthor={Author},     
    pdfsubject={Subject},   
    pdfcreator={Creator},   
    pdfproducer={Producer}, 
    pdfkeywords={keyword1} {key2} {key3}, 
    pdfnewwindow=true,      
    colorlinks=true,       
    linkcolor=red,          
    citecolor=blue,        
    filecolor=magenta,      
    urlcolor=red           
}

\definecolor{light-gray}{gray}{0.7}
\def\red#1 {\textcolor{red}{#1}\ }   
\def\green#1 {\textcolor{green}{#1}\ }   
\def\cyan#1 {\textcolor{cyan}{#1}\ }   
\def\ggray#1 {\textcolor{light-gray}{#1}\ }

\def\gs{\mathrel{\raise0.35ex\hbox{$\scriptstyle >$}\kern-0.6em \lower0.40ex\hbox{{$\scriptstyle \sim$}}}}
\def\ls{\mathrel{\raise0.35ex\hbox{$\scriptstyle <$}\kern-0.6em \lower0.40ex\hbox{{$\scriptstyle \sim$}}}}
\newcommand{\Msolar}{\mbox{$M_{\odot}\,$}}
\newcommand{\Lsolar}{\mbox{$L_{\odot}\,$}}

\newcommand{\arcsecs}{\mbox{$^{\prime\prime}$}}

\shorttitle{Star formation relations and CO SLEDs across the $J$-ladder and redshift}
\shortauthors{Greve et al.}

\begin{document}


\title{Star formation relations and CO Spectral Line Energy Distributions across the $J$-ladder and redshift }


\author{
 T.~R.~Greve\altaffilmark{1},
I.~Leonidaki\altaffilmark{2},		
E.~M.~Xilouris\altaffilmark{2},
A.~Wei\ss\altaffilmark{3},
Z.-Y.~Zhang\altaffilmark{4,5},
P.~van der Werf\altaffilmark{6},
S.~Aalto\altaffilmark{7},
L.~Armus\altaffilmark{8},
T.~D\'iaz-Santos\altaffilmark{8},
A.S.~Evans\altaffilmark{9,10},
J.~Fischer\altaffilmark{11},
Y.~Gao\altaffilmark{12},
E.~Gonz\'alez-Alfonso\altaffilmark{13},
A.~Harris\altaffilmark{14},
C.~Henkel\altaffilmark{3},
R.~Meijerink\altaffilmark{6,15},
D.~A.~Naylor\altaffilmark{16}
H.~A.~Smith\altaffilmark{17}
M.~Spaans\altaffilmark{15}
G.~J.~Stacey\altaffilmark{18}
S.~Veilleux\altaffilmark{14}
F.~Walter\altaffilmark{19}
}

\altaffiltext{1}{Department of Physics and Astronomy, University College London, Gower Street, London WC1E 6BT, UK}
\altaffiltext{2}{Institute for Astronomy, Astrophysics, Space Applications \& Remote Sensing, National Observatory of Athens, GR-15236 Penteli, Greece}
\altaffiltext{3}{Max-Planck-Institut fur Radioastronomie, Auf dem H\"ugel 69, D-53121 Bonn, Germany}
\altaffiltext{4}{UK Astronomy Technology Centre, Science and Technology Facilities Council, Royal Observatory, Blackford Hill, Edinburgh EH9 3HJ, UK}
\altaffiltext{5}{European Southern Observatory, Karl Schwarzschild Stra{\ss}e 2, 85748 Garching, Germany}
\altaffiltext{6}{Leiden Observatory, Leiden University, PO Box 9513, NL-2300 RA Leiden, the Netherlands}
\altaffiltext{7}{Department of Earth and Space Sciences, Chalmers University of Technology, Onsala Observatory, 43994 Onsala, Sweden}
\altaffiltext{8}{Spitzer Science Center, California Institute of Technology, MS 220-6, Pasadena, CA 91125, USA}
\altaffiltext{9}{Astronomy Department, University of Virginia Charlottesville, VA 22904, USA}
\altaffiltext{10}{National Radio Astronomy Observatory, 520 Edgemont Road, Charlottesville, VA 22903, USA}
\altaffiltext{11}{Naval Research Laboratory, Remote Sensing Division, 4555 Overlook Ave SW, Washington, DC 20375, USA}
\altaffiltext{12}{Purple Mountain Observatory, Chinese Academy of Sciences, 2 West Beijing Road, Nanjing 210008, China}
\altaffiltext{13}{Universidad de Alcal´a de Henares, Departamento de F\'ısica, Campus Universitario, E-28871 Alcal\'a de Henares, Madrid, Spain}
\altaffiltext{14}{Department of Astronomy, University of Maryland, College Park, MD 20742, USA}
\altaffiltext{15}{Kapteyn Astronomical Institute, University of Groningen, P.O. Box 800, 9700 AV Groningen, the Netherlands}
\altaffiltext{16}{Institute for Space Imaging Science, Department of Physics and Astronomy, University of Lethbridge, Lethbridge, AB T1K 3M4, Canada}
\altaffiltext{17}{Harvard–Smithsonian Center for Astrophysics, 60 Garden Street, Cambridge, MA 02138, USA}
\altaffiltext{18}{Department of Astronomy, Cornell University, Ithaca, NY 1485, USA}
\altaffiltext{19}{Max-Planck-Institut fur Astronomie, K\"onigstuhl 17, D-691117 Heidelberg, Germany}
\email{t.greve@ucl.ac.uk}

\begin{abstract}
We present FIR[$50-300\,{\rm \mu m}$]$-$CO luminosity relations (i.e.,
$\log L_{\rm FIR} = \alpha \log L'_{\rm CO} + \beta$) for the full CO rotational
ladder from $J=1-0$ up to $J=13-12$ for a sample of 62 local ($z \le 0.1$) (Ultra) Luminous Infrared Galaxies (LIRGs; $L_{\rm IR[8-1000\,\mu m]} >
10^{11}\,\Lsolar$) using data from {\it Herschel} SPIRE-FTS and ground-based
telescopes. We extend our sample to high redshifts ($z > 1$) by including 35
(sub)-millimeter selected dusty star forming galaxies from the literature with
robust CO observations, and sufficiently well-sampled FIR/sub-millimeter
spectral energy distributions (SEDs) so that accurate FIR luminosities can be
deduced. The addition of luminous starbursts at high redshifts enlarge the range
of the FIR$-$CO luminosity relations towards the high-IR-luminosity end while
also significantly increasing the small amount of mid-$J$/high-$J$ CO line data
($J=5-4$ and higher) that was available prior to {\it Herschel}.
This new data-set (both in terms of IR luminosity and $J$-ladder) reveals linear
FIR$-$CO luminosity relations (i.e., $\alpha\simeq 1$) for $J=1-0$ up to
$J=5-4$, with a nearly constant normalization ($\beta \sim 2$). In the
simplest physical scenario this is expected from the (also) linear
FIR$-$(molecular line) relations recently found for the dense gas tracer lines
(HCN and CS), as long as the dense gas mass fraction does not vary strongly
within our (merger/starburst)-dominated sample. However from $J=6-5$ and up to
the $J=13-12$ transition we find an increasingly sub-linear slope and higher
normalization constant with increasing $J$. We argue that these are caused by a
warm ($\sim 100\,\rm{K}$) and dense ($>10^{4}\,\rm{cm^{-3}}$) gas component
whose thermal state is unlikely to be maintained by star formation powered
far-UV radiation fields (and thus is no longer directly tied to the star
formation rate). We suggest that mechanical heating (e.g., supernova
driven turbulence and shocks), and not cosmic rays, is the more likely source of
energy for this component. The global CO spectral line energy distributions
(SLEDs), which remain highly excited from $J=6-5$ up to $J=13-12$, are found to
be a generic feature of the (U)LIRGs in our sample, and further support the
presence of this gas component.
\end{abstract}





\section{Introduction}
Early empirical correlations between the preponderance of young stars and gas in
galaxies (e.g., \citet{sanduleak1969}) confirmed -- in a qualitative sense --
the simple power-law dependence between star formation rate surface density
($\Sigma_{\rm SFR}$) and gas surface density ($\Sigma_{\rm gas}$) first
suggested by \citet{schmidt1959} who found $\Sigma_{\rm SFR} \propto \Sigma_{\rm
gas}^2$ for H{\sc i} gas. Once the H$_2$ component as traced by CO lines was
identified in galaxies, the gas surface density could be related to both H\,{\sc
i} and H$_2$, i.e., $\Sigma_{\rm gas}=\Sigma_{\rm HI}+\Sigma_{\rm H_2}$
\citep{kennicutt1989}. In a seminal paper, \citet{kennicutt1998} established
this relation, hereafter called the Schmidt-Kennicutt (S-K) relation, to be:
$\Sigma_{\rm SFR} \propto \Sigma_{\rm gas}^{1.4}$, averaged over entire galaxy
disks. Further studies by \citet{wong2002} and \citet{schruba2011} found a
nearly linear S-K relation for the molecular gas on kpc scales (see also
\citet{bigiel2008} and \citet{leroy2008,leroy2013}), with the SFR surface
density having a much closer correspondence with the molecular gas surface
density -- reflecting the well-established fact that stars form out of molecular
rather than atomic gas.  Much theoretical effort has gone into obtaining the
exponents and normalization of this relation as unique outcomes of various
physical processes occurring in star forming galaxies, with various models
capable of yielding (S-K)-type relations (e.g.,
\citet{dopita1994,gerritsen1997,wong2002,elmegreen2002}). It became evident
that, while no deterministic microphysics of the interstellar medium (ISM) and
star formation (SF) can be linked to a given S-K relation, the high-density gas
component ($n\geq 10^{4}\,{\rm cm^{-3}}$) plays a crucial role in ultimately
anchoring such relations to the star formation taking place deep inside
supersonically turbulent molecular clouds in disks.

The S-K relations for high-density gas are particularly challenging to establish
since determining the dense gas mass fraction within a galaxy requires
observations of CO from $J=1-0$ (a total molecular gas mass tracer) up to at
least $J=3-2$ along with the much fainter lines of {\it bona fide} dense gas
tracers like CS and heavy-rotor molecules such as HCN. A multi-component
analysis of such CO, HCN, and CS spectral line energy distributions
(SLEDs) can then yield dense gas masses, $M_{\rm dense}(n\geq 10^{4}\,{\rm
cm^{-3}})$ (e.g., \citet{mao2000,greve2009}). However, to do so for a large
number of galaxies in order to obtain even a surface-integrated SFR$-M_{\rm
dense}$ S-K relation has been prohibitively expensive in telescope time. At high
redshifts the situation is made worse due to a lack in sensitivity and angular
resolution. Nonetheless, pioneering efforts have been made at discerning
$\Sigma_{\rm SFR} = A \Sigma_{\rm gas}^{N}$ at high redshifts using H$\alpha$
maps obtained with integral field unit cameras, and high-resolution
interferometric CO ($J=1-0$ to $3-2$) observations of massive star forming
galaxies at $z\sim 1-3$ \citep{genzel2010,tacconi2013,freundlich2013}.
Obviously, this situation will now improve dramatically with the advent of the
Atacama Large Millimeter/Sub-millimeter Array (ALMA).

With the dense gas mass fraction distribution currently inaccessible for any
statistically significant number of galaxies one must fall back to the
integrated (S-K)-proxy relations: $L_{\rm IR}-L_{\rm line}$ (where $L_{\rm
line}$ is the line luminosity of a dense gas tracer and $L_{\rm IR}$ a linear
proxy of SFR), and then invoke theoretically determined links to an underlying
S-K relation \citep{krumholz2007,narayanan2008}. HCN($1-0$) observations of
statistically significant samples of local IR luminous galaxies (LIRGs) and
normal spiral galaxies yielded the first of such (S-K)-proxy relations using gas
tracers other than CO lines \citep{solomon1992}, finding the IR$-$HCN relation
to be linear and with much less scatter than the previously determined IR$-$CO
low-$J$ relations.  This was interpreted as HCN($1-0$), with its high critical
density ($\sim 10^5\,{\rm cm^{-3}}$), being a more direct tracer of a dense,
star forming gas component with a nearly constant underlying star formation
efficiency (SFE) \citep{gao2004a,gao2004b}. Furthermore, with the tight, linear
IR$-$HCN relation extending down to individual Galactic molecular clouds where
$L_{\rm IR} \gs 10^{4.5}\,{\rm \Lsolar}$, thus covering over $\sim 8$ orders of
magnitude in luminosity, its origin could be attributed to the existence of
fundamental `units' of cluster star formation \citep{wu2005}. This view is now
further supported by the linear $L_{\rm IR}-L_{\rm line}$ relations found also
for the HCN($4-3$) and CS($7-6$) lines \citep{zhang2014}, which for CS($7-6$)
also extends (linearly) down to Galactic cores \citep{wu2010}. Some contentious
points do remain however, especially towards the high-$L_{\rm IR}$ end which is
dominated by mergers/starbursts where a slightly super-linear IR$-$HCN relation
has been claimed and argued to be due mostly to an increase in the dense gas SFE
in such galaxies \citep{riechers2007,GC-B-2008}. 

In this paper we present the first FIR$-$CO luminosity relations and the
corresponding global CO SLEDs that extend above $J_{\rm up} = 4$ and up to
$J_{\rm up} = 13$ using {\it Herschel} SPIRE-FTS data for local (U)LIRGs. The
FIR$-$CO relations and CO SLEDs presented in this work (from $J=1-0$ up to
$J=13-12$), besides a significant extension of the $J$-ladder, benefit also from
the inclusion of (U)LIRGs from the low- and the high-$z$ Universe. This robustly
extends the sample towards the important high-$L_{\rm FIR}$ end (as numerous
galaxies with ULIRG-like, or higher, luminosities have been found in the
high-$z$ Universe) where very different conditions may prevail for the molecular
gas, possibly leaving an imprint on the FIR$-$CO relations and the CO SLEDs. Our
new high-$J$ CO line data-set is uniquely sensitive to such an imprint since
these lines need both high densities ($n_{\rm crit}\sim (10^{4}-7\times
10^{5})\,{\rm cm^{-3}}$) {\it and} (in most circumstances) high temperatures
($E_{J}/k_{\rm B} \sim (55-500)\,{\rm K}$) to be significantly excited. The
high-density and warm gas necessary for exciting them is the most difficult
phase to maintain energetically in appreciable quantities in galaxies.  However,
it is one that would leave no easily discernible signature in the low-$J$ CO and
low-/mid-$J$ SLEDs of heavy rotor molecular lines (e.g., HCN, CS) that typically
have been available for (U)LIRGs up to now.  Throughout, we adopt a flat
cosmology with $\Omega_{\rm M} = 0.315, \Omega_{\Lambda} = 0.685$, and $h =
0.67$ \citep{planck2013}.

\section{Galaxy samples and data}\label{section:samples-and-data}
For the purposes of this work we first compiled high-$J$ ($J=4-3$ up to
$J=13-12$ line data from the {\it Herschel} Comprehensive (U)LIRG Emission
Survey (HerCULES, \citet{vanderwerf2010}) -- an open time key program on the ESA
{\it Herschel} Space Observatory\footnote{Herschel is an ESA space observatory
with science instruments provided by European-led Principal Investigator
consortia and with important participation from NASA.} \citep{pilbratt2010}
which measured CO $J=4-3$ to $J=13-12$ for 29 local ($z<0.1$) (U)LIRGs using the
Fourier Transform Spectrometer (FTS) of the SPIRE instrument\footnote{SPIRE has
been developed by a consortium of institutes led by Cardiff Univ. (UK) and
including: Univ. Lethbridge (Canada); NAOC (China); CEA, LAM (France); IFSI,
Univ. Padua (Italy); IAC (Spain); Stockholm Observatory (Sweden); Imperial
College London, RAL, UCL-MSSL, UKATC, Univ.  Sussex (UK); and Caltech, JPL,
NHSC, Univ. Colorado (USA). This development has been supported by national
funding agencies: CSA (Canada); NAOC (China); CEA, CNES, CNRS (France); ASI
(Italy);MCINN (Spain); SNSB (Sweden); STFC, UKSA (UK); and NASA (USA).}
\citep{griffin2010}. The HerCULES sources were selected from the $60\,{\rm \mu
m}$ flux-limited IRAS Revised Bright Galaxy Sample ($f_{\rm 60\mu m} >
5.24\,{\rm Jy}$; \citet{sanders2003}) with separate flux cuts applied to ULIRGs
and LIRGs ($f_{\rm 60\mu m} > 11.65\,{\rm Jy}$ and $>16.4\,{\rm Jy}$,
respectively). A detailed description of the SPIRE-FTS observations, calibration
modes, extraction of CO line fluxes, and final line luminosities, are given in a
dedicated paper (Rosenberg et al., in prep.).  Briefly, the high spectral
resolution mode was used with a resolution of $1.2\,{\rm GHz}$ over both
observing bands.  A reference measurement was used to subtract the emission from
the sky, telescope, and instrument.  The spectra were reduced using the Herschel
Interactive Processing Environment (HIPE), ver.\ 9.0.  At the time of writing,
fully reduced SPIRE-FTS CO spectra were available for only 26 sources, and of
these three had extended, multi-component morphologies and were discarded.
Since the SPIRE-FTS beam ranges from $\sim 16\arcsecs$ to $\sim 42\arcsecs$
(FWHM) across the bandpass \citep{makiwa2013}, it is essential to perform a beam
correction in cases where the sources are extended with respect to the beam. All
spectra (and thus CO line fluxes) were scaled to a common spatial resolution of
$\sim 42\arcsecs$ using LABOCA $870\,\mu m$ or SABOCA $350\,{\rm \mu m}$ maps
(see Rosenberg et al., in prep. for details).  Obviously, this assumes that the
corrections are perfectly mono-chromatic in the FIR and sub-millimeter (sub-mm)
regime, which is a good assumption to within $\ls 20\%$ \citep{galametz2013}.
For the HerCULES sources, which are all (U)LIRGs and thus nearly all relatively
compact, and well within the beam sizes of the CO observations, this correction
was minor.  For very extended sources, however, this correction is crucial, and
failing to apply it can skew the observed FIR$-$CO relation (i.e.,
\citet{bussmann2008}, \citet{juneau2009}, and see discussion in
\citet{zhang2014}).
 
\begin{figure}[h]
\includegraphics[scale=0.49]{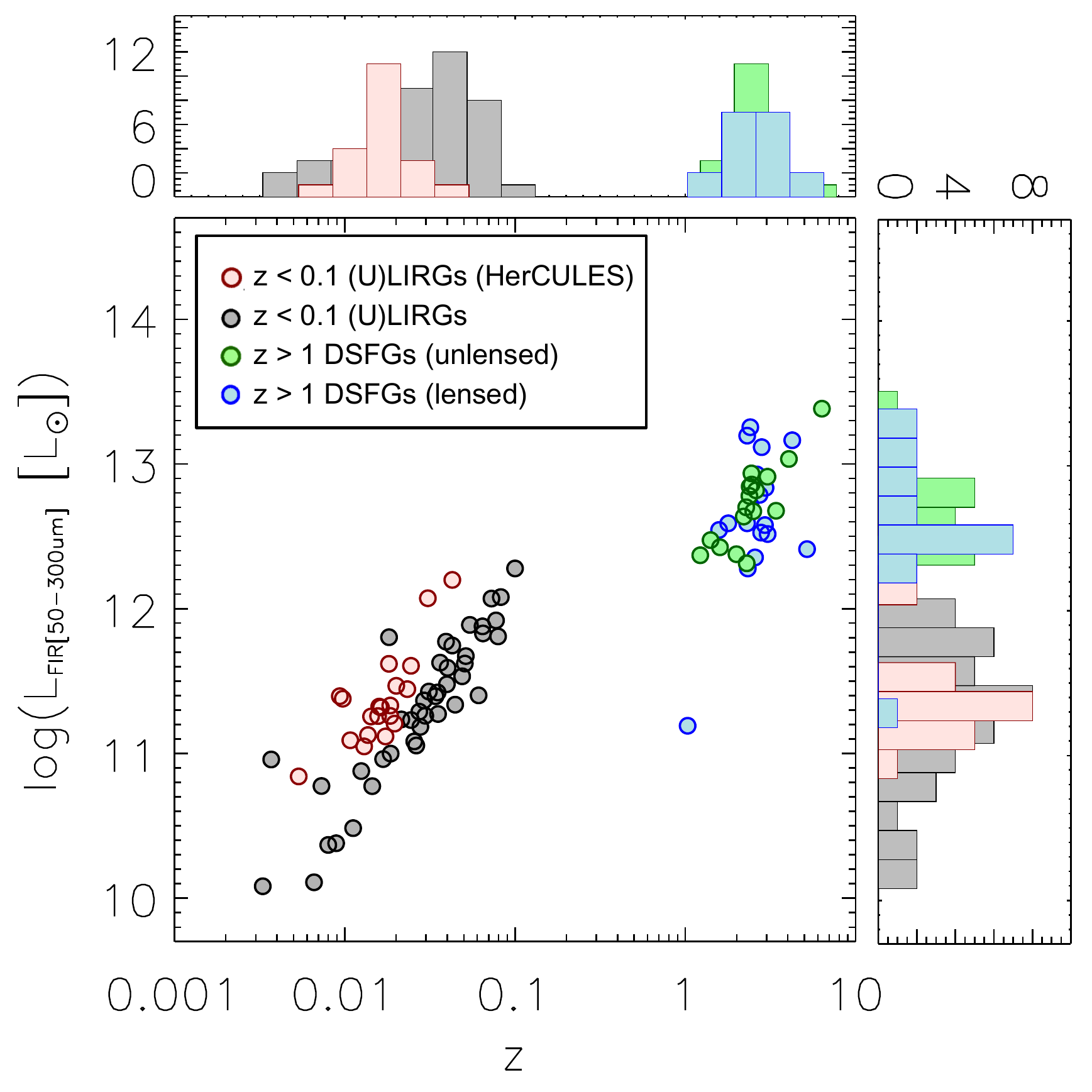}
\caption{The logarithm of the FIR ($50-300\,{\rm \mu m}$) luminosity vs.\
redshift for the galaxy samples considered in this paper (after AGN-dominated
systems have been removed), along with histograms of the FIR luminosity and
redshift distributions (top and right inserts, respectively). The local
($z<0.1$) sources include sub-sets of the (U)LIRG samples from HerCULES (20
sources, red symbols) and \citet{papadopoulos2012} (42 sources, grey symbols)
The high-$z$ ($z>1$) sources are unlensed, or weakly lensed, DSFGs (16 sources,
green symbols) and strongly lensed DSFGs (19 sources, blue symbols) uncovered
from various (sub)-mm surveys (see \S~\ref{section:samples-and-data}). All FIR
luminosities have been been corrected for lensing using the magnification
factors in Table \ref{table:high-z-sample}.  
}
\label{figure:fig0}
\end{figure}

We also included ground-based CO line data presented by \citet{papadopoulos2012}
for a sample of 45 local (U)LIRGs\footnote{The full sample in
\citet{papadopoulos2012} consisted of 70 (U)LIRGs, but 25 of those lacked
adequate continuum FIR and/or sub-mm data and were discarded.} from the {\it
IRAS} RBGS. These data consisted of low-$J$ CO transitions, i.e., $J=1-0$ (all
45 sources), $2-1$ (17), $3-2$ (44), as well as $J=4-3$ (3) and $6-5$ (12)
observations. This allowed us to both fill-in the $J=1-0$, $2-1$, $3-2$
transitions for the 11 HerCULES sources that overlapped with this sample (except
for one source which did not have $J=2-1$ measurement), and bring in additional
CO low-$J$ and $J=4-3/6-5$ lines (the only mid-/high-$J$ CO lines accessible
with the ground-based telescopes used -- see \citet{papadopoulos2012} for
details) to the sample. We stress that the CO line fluxes given in
\citet{papadopoulos2012} are {\it total} line fluxes, and so no additional beam
correction is required for these sources.

Of our sample of 68 local (U)LIRGs (listed in Table \ref{table:low-z-sample}),
30 sources (20+10 from HerCULES and  \citet{papadopoulos2012} sub-samples,
respectively) are also part of The Great Observatories All-Sky LIRG Survey
(GOALS; \citet{stierwalt2013}). To weed out active galactic nuclei (AGN), the
sample was cross-correlated against estimates of the AGN contribution to the
bolometric luminosity based on several MIR diagnostics such as the equivalent
width of the $6.2\,{\rm \mu m}$ PAH feature, the [Ne\,{\sc v}]$/$[Ne\,{\sc ii}]
and [O\,{\sc iv}]$/$[Ne\,{\sc ii}] emission line ratios as well as
$30$-to-$15\,{\rm \mu m}$ continuum flux ratios \citep{veilleux2009,petric2011}.
Only six sources (indicate by a $*$ in Table \ref{table:low-z-sample}) were
found to have an AGN contribution $> 30\%$ and were omitted from our analysis
(although, including them in our analysis did not alter the findings of this
paper). 

The FIR/sub-mm continuum data were obtained from a number of studies (see
\citet{papadopoulos2012} and references therein) as well as from the NASA/IPAC
Extragalactic Database (NED). All the $850\,{\rm \mu m}$ and $1.2\,{\rm mm}$
available fluxes were corrected for CO $J=3-2$, and $2-1$ line contamination
($<20\%$), respectively. We also corrected for any non-thermal radio continuum
contributions whenever radio data were available, allowing for a power-law
extrapolation to the sub-mm wavelengths. The FIR ($50-300\,{\rm \mu m}$)
luminosities derived from the continuum data (see \S~\ref{section:analysis} for
details) span the range $\sim 10^{10-12}\,{\rm \Lsolar}$ (Fig.\
\ref{figure:fig0}). The two samples are well matched in luminosity, although no
HerCULES sources are found at $\ls 10^{10.8}\,{\rm \Lsolar}$.  The fact that the
more luminous sources tend to have higher redshifts merely reflects the
flux-limited selection of the two samples.

\bigskip

High-redshift dusty star forming galaxies (DSFGs\footnote{In this paper we take
DSFGs to be synonymous with highly dust-enshrouded major merger starbursts
selected at sub-mm/mm wavelengths (also often referred to as (sub)-millimeter
selected galaxies, i.e., SMGs)}) are thought to resemble the local (U)LIRG
population and most have $L_{\rm IR}\gs 10^{12}\,{\rm \Lsolar}$.  Moreover,
typically multiple high-$J$ CO lines and FIR/(sub)-mm continuum observations are
available for them. These were the main reasons for including them in our
analysis. In order to achieve the best possible uniformity, a meticulous
compilation of the aforementioned observations (CO line and continuum
observations) for all published DSFGs was extracted from the literature (guided
by major review papers by \citet{solomon2005} and \citet{carilli2013}). Sources
with clear signs of AGN (e.g., from optical spectroscopy showing strong
Ly$\alpha$, C\,{\sc iv}, and C\,{\sc iii} emission lines and a power-law
continuum, or radio-loudness) were not included in our sample. In cases where
multiple observations of the same CO transition existed, we adopted the weighted
mean of the velocity-integrated line flux after discarding any outliers and
measurements with low signal-to-noise.  Many of the high-$z$ CO detections are
of strongly lensed DSFGs, which we here take to mean a gravitational
magnification factor ($\mu$) $>1$, and in those cases we adopted the best
estimates of $\mu$ available at the time of writing (e.g.,
\citet{swinbank2010,aravena2013,bussmann2013}). A total of 74 DSFGs constituted
our initial high-$z$ sample. For 39 (53\%) of the DSFGs, however, we were unable
to put reliable constraints on their FIR luminosities (see
\S~\ref{section:analysis}), and these were therefore discarded for the analysis
presented in this paper. This left us with a final sample of 35 high-$z$ sources
(listed in Table \ref{table:high-z-sample}), spanning the redshift range
$z=1.0-6.3$ with a median redshift of $z\simeq 2.4$ (see also Fig.\
\ref{figure:fig0}). The lensed DSFGs (19 in total), {\it after magnification
correction}, span a similar range in FIR luminosity as the non-lensed DSFGs
($\sim 10^{12-14}\,{\rm \Lsolar}$, see Fig.\ \ref{figure:fig0}), which is about
an order of magnitude higher than that of the local samples. The only exception
is SMM\,J163555.2$+$661150 ($z=1.03$), which has an intrinsic luminosity similar
to that of local LIRGs \citep{knudsen2009}.  Within the high-$z$ samples, we see
no strong dependence of FIR luminosity on redshift, which is due to the
well-known flat selection function at sub-mm wavelengths for $z\gs 1$
\citep{blain1993}. Finally, we stress that while the DSFGs, as a sample, cover
all CO transitions from $J=1-0$ to $J=10-9$, no individual galaxy has continuous
coverage across this transition range.

%
\setlength\LTleft{-30pt}            
\setlength\LTright{-30pt}           
\setlength\LTcapwidth{385pt}           
\begin{longtable*}[t]{lccc}
\caption{The sample of 68 local ($z < 0.1$) (U)LIRGs used in this paper. The
first 23 sources listed below (and {\it not} listed in italics) were observed by
{\it Herschel}/SPIRE-FTS as part of the HerCULES program
(\S\,\ref{section:samples-and-data}).  Sources indicated by a $*$ were found to
have significant AGN contribution ($>30\%$ of the bolometric luminosity) and
were not included in our final analysis.}\\
\hline
ID                   	                              & $z$     & $\log ( L_{\rm FIR[50-300\,\mu m]} / \Lsolar )$ & $\log (L_{\rm IR[8-1000\,\mu m]} / \Lsolar )$  \\
\hline
\hline
IRAS\,00085$-$1223 (NGC\,34)                          &0.0196   &   11.21                     &11.47\\
IRAS\,00506$+$7248 (MCG$+$12-02-001)                  &0.0157   &   11.26                     &11.53\\
IRAS\,01053$-$1746 (IC\,1623)                         &0.0201   &   11.47                     &11.74\\
IRAS\,04315$-$0840 (NGC\,1614)                        &0.0159   &   11.32                     &11.59\\
IRAS\,05189$-$2524$*$                                 &0.0426   &   11.73                     &12.12\\
IRAS\,08354$+$2555 (NGC\,2623)                        &0.0185   &   11.33                     &11.60\\
IRAS\,10257$-$4339 (NGC\,3256)                        &0.0094   &   11.40                     &11.60\\
IRAS\,11506$-$3851 (ESO\,320$-$G030)                  &0.0108   &   11.09                     &11.30\\
IRAS\,12540$+$5708 (Mrk\,231)$*$                      &0.0422   &   12.14                     &12.56\\
IRAS\,13120$-$5453 (WKK\,2031)                        &0.0308   &   12.07                     &12.34\\
IRAS\,13183$+$3423 (Arp\,193)                         &0.0233   &   11.44                     &11.68\\
IRAS\,13229$-$2934 (NGC\,5135)                        &0.0137   &   11.13                     &11.33\\
IRAS\,13242$-$5713 (ESO\,173$-$G015)                  &0.0097   &   11.38                     &11.65\\
IRAS\,13428$+$5608 (Mrk\,273)$*$                      &0.0378   &   11.91                     &12.17\\
IRAS\,16504$+$0228 (NGC\,6240)                        &0.0245   &   11.61                     &11.87\\
IRAS\,15107$+$0724 (Zw\,049.057)                      &0.0130   &   11.05                     &11.28\\
IRAS\,17208$-$0014                                    &0.0428   &   12.20                     &12.47\\
IRAS\,18093$-$5744 (IC\,4687)                         &0.0173   &   11.12                     &11.39\\
IRAS\,18293$-$3413                                    &0.0182   &   11.62                     &11.84\\
IRAS\,23007$+$0836 (NGC\,7469)                        &0.0163   &   11.32                     &11.60\\
IRAS\,23134$-$4251 (NGC\,7552)                        &0.0054   &   10.84                     &11.05\\
IRAS\,23488$+$2018 (Mrk\,331)                         &0.0185   &   11.26                     &11.53\\
IRAS\,23488$+$1949 (NGC\,7771)                        &0.0143   &   11.26                     &11.43\\
{\it IRAS\,00057$+$4021}                              &0.0445   &   11.34                     &11.60\\
{\it IRAS\,00509$+$1225}                              &0.0611   &   11.40                     &11.67\\
{\it IRAS\,01077$-$1707}                              &0.0351   &   11.42                     &11.69\\
{\it IRAS\,01418$+$1651}                              &0.0274   &   11.29                     &11.56\\
{\it IRAS\,02114$+$0456}                              &0.0297   &   11.26                     &11.43\\
{\it IRAS\,02401$-$0013}                              &0.0037   &   10.96                     &11.23\\
{\it IRAS\,02483$+$4302}                              &0.0514   &   11.67                     &11.85\\
{\it IRAS\,02512$+$1446}                              &0.0312   &   11.43                     &11.70\\
{\it IRAS\,03359$+$1523}                              &0.0353   &   11.27                     &11.45\\
{\it IRAS\,04232$+$1436}                              &0.0795   &   11.81                     &12.08\\
{\it IRAS\,05083$+$7936}                              &0.0543   &   11.88                     &12.06\\
{\it IRAS\,08572$+$3915}$*$                           &0.0582   &   11.73                     &12.11\\
{\it IRAS\,09126$+$4432}                              &0.0398   &   11.48                     &11.65\\
{\it IRAS\,09320$+$6134}                              &0.0393   &   11.77                     &11.95\\
{\it IRAS\,09586$+$1600}                              &0.0080   &   10.37                     &10.64\\
{\it IRAS\,10035$-$4852}                              &0.0648   &   11.83                     &12.10\\
{\it IRAS\,10039$-$3338}$*$                           &0.0341   &   11.47                     &11.74\\
{\it IRAS\,10173$+$0828}                              &0.0489   &   11.53                     &11.80\\
{\it IRAS\,10356$+$5345}                              &0.0033   &   10.08                     &10.35\\
{\it IRAS\,10565$+$2448}                              &0.0428   &   11.75                     &12.01\\
{\it IRAS\,11231$+$1456}                              &0.0341   &   11.40                     &11.57\\
{\it IRAS\,12001$+$0215}                              &0.0066   &   10.11                     &10.29\\
{\it IRAS\,12112$+$0305}                              &0.0727   &   12.07                     &12.34\\
{\it IRAS\,12224$-$0624}                              &0.0263   &   11.05                     &11.23\\
{\it IRAS\,12243$-$0036}                              &0.0073   &   10.77                     &11.04\\
{\it IRAS\,13001$-$2339}                              &0.0215   &   11.24                     &11.41\\
{\it IRAS\,13102$+$1251}                              &0.0112   &   10.48                     &10.66\\
{\it IRAS\,13188$+$0036}                              &0.0186   &   11.00                     &11.17\\
{\it IRAS\,13362$+$4831}                              &0.0278   &   11.18                     &11.45\\
{\it IRAS\,13470$+$3530}                              &0.0168   &   10.96                     &11.10\\
{\it IRAS\,13564$+$3741}                              &0.0125   &   10.88                     &11.05\\
{\it IRAS\,14003$+$3245}                              &0.0145   &   10.77                     &10.95\\
{\it IRAS\,14178$+$4927}                              &0.0256   &   11.08                     &11.35\\
{\it IRAS\,14348$-$1447}                              &0.0825   &   12.08                     &12.42\\
{\it IRAS\,15163$+$4255}                              &0.0402   &   11.59                     &11.94\\
{\it IRAS\,15243$+$4150}                              &0.0089   &   10.38                     &10.65\\
{\it IRAS\,15327$+$2340}                              &0.0182   &   11.80                     &11.98\\
{\it IRAS\,15437$+$0234}$*$                           &0.0128   &   10.84                     &11.01\\
{\it IRAS\,16104$+$5235}                              &0.0292   &   11.37                     &11.63\\
{\it IRAS\,16284$+$0411}                              &0.0245   &   11.23                     &11.40\\
{\it IRAS\,17132$+$5313}                              &0.0507   &   11.62                     &11.89\\
{\it IRAS\,19458$+$0944}                              &0.1000   &   12.28                     &12.45\\
{\it IRAS\,20550$+$1656}                              &0.0363   &   11.63                     &11.97\\
{\it IRAS\,22491$-$1808}                              &0.0773   &   11.92                     &12.19\\
{\it IRAS\,23365$+$3604}                              &0.0644   &   11.88                     &12.15\\
\hline
                                                      &         &                             &      \\
                                                      &         &                             &      \\
\label{table:low-z-sample}
\end{longtable*}

%
\begin{table*}[t]
\begin{threeparttable}
\caption{The high-$z$ DSFGs samples utilized in this paper, consisting of
35 sources in total of which 19 are strongly lensed, i.e., gravitational
magnification factor $\mu > 1$ (bottom 19, shown in italics).  The listed FIR
($50-300\,{\rm \mu m}$) and IR ($8-1000\,{\rm \mu m}$) luminosities have not
been corrected for gravitational lensing but we give the most up to date
estimates of the magnification ($\mu$) factor (from the literature) needed to
perform this correction along with the appropriate references for each source.
For completeness and for cross-comparison, we also give alternative, but now
most likely outdated, magnification estimates in parentheses.
}
\begin{tabular}{lcccll}
\hline
ID                   	                              & $z$     & $\log ( L_{\rm FIR} / \Lsolar )$ & $\log (L_{\rm IR} / \Lsolar )$ & $\mu$ & ref.\\

\hline
\hline
SMM\,J021725$-$045934 (SXDF\,11)                                 &  2.2920    & 12.30                         & 12.51	& 1.0  & [1,2]\\ 
SMM\,J030227.73$+$000653.3                                       &  1.4060    & 12.45                         & 12.72	& 1.0  & [1]\\
SMM\,J105151.69$+$572636.0 (Lock850.16)                          &  1.5973    & 12.40                         & 12.67	& 1.0  & [1]\\
SMM\,J105227.58$+$572512.4 (LE\,1100.16)                         &  2.4432    & 12.92                         & 13.26	& 1.0  & [1]\\
SMM\,J105230.73$+$572209.5 (LE\,1100.05)                         &  2.6011    & 12.80                         & 13.08	& 1.0  & [1,3]\\
SMM\,J105238.30$+$572435.8 (LE\,1100.08)                         &  3.0360    & 12.90                         & 13.17	& 1.0  & [3]\\
SMM\,J123549.44$+$621536.8 (AzGN\,15, HDF\,76)                   &  2.2020    & 12.62                         & 12.89	& 1.0  & [1,4]\\
SMM\,J123600.16$+$621047.3                                       &  1.9941    & 12.36                         & 12.53	& 1.0  & [1,3]\\
SMM\,J123606.85$+$621047.2                                       &  2.5054    & 12.66                         & 12.83	& 1.0  & [1]\\
SMM\,J123634.51$+$621240.9 (GN\,26, HDF\,169)                    &  1.2224    & 12.35                         & 12.61	& 1.0  & [5,6]\\
SMM\,J123711.86$+$622212.6 (GN\,20, AzGN\,01)                    &  4.0554    & 13.03                         & 13.23	& 1.0  & [7,8]\\
SMM\,J131201.17$+$424208.1                                       &  3.4078    & 12.67                         & 12.94	& 1.0  & [3,6,9]\\
SMM\,J163631.47$+$405546.9 (N2\,850.13)                          &  2.2767    & 12.69                         & 12.96	& 1.0  & [1,3]\\
SMM\,J163658.19$+$410523.8 (N2\,850.02)                          &  2.4546    & 12.84                         & 13.11	& 1.0  & [3,4,10]\\
SMM\,J163650.43$+$405734.5 (N2\,850.04)                          &  2.3853    & 12.83                         & 13.10	& 1.0  & [1,10,11]\\
SMM\,J163706.51$+$405313.8 (N2\,1200.17)                         &  2.3774    & 12.77                         & 12.96	& 1.0  & [1,3]\\
{\it 1HERMESS250\,J022016.5$−$060143 (HXMM01)}                   &  2.3074    & 13.17                         & 13.37   & $1.5\pm 0.3$                                 & [12]\\
{\it SMM\,J02399$-$0136}                                         &  2.8076    & 13.08                         & 13.43	& $2.38\pm 0.08$ (2.45)                        & [13,14]\\
{\it SPT-S\,J053816$-$5030.8}                                    &  2.7818    & 12.49                         & 12.69   & $20\pm 4$                                    & [15,16]\\
{\it HATLAS\,J084933.4$+$021443-T}                               &  2.4090    & 12.98                         & 13.19	& $2.8\pm 0.2$ ($1.5\pm 0.2$)                  & [17,18]\\
{\it HATLAS\,J084933.4$+$021443-W}                               &  2.4068    & 13.24                         & 13.51	& 1.0                                          & [17]\\
{\it H-ATLASJ090302.9$-$014128-17b (SDP.17b)}                    &  2.3051    & 12.01                         & 12.21	& $4.9\pm 0.7$ ($18\pm 8$)                     & [19,20,21]\\
{\it H-ATLASJ090311.6$$+$$003906 (SDP.81)}                       &  3.0425    & 12.15                         & 12.35	& $11.1\pm 1.1$ ($14\pm 4$, 18-31)             & [18,20,22]\\
{\it H-ATLASJ090740.0$-$004200 (SDP.9)}                          &  1.5770    & 13.47                         & 13.67	& $8.8\pm 2.2$                                 & [18,20]\\
{\it H-ATLASJ091043.1$-$000322 (SDP.11)}                         &  1.7860    & 13.61                         & 13.88	& $10.9\pm 1.3$                                & [18,20]\\
{\it H-ATLASJ091305.0$-$005343 (SDP.130)}                        &  2.6256    & 12.46                         & 12.66	& $2.1\pm 0.3$ (5-7, $10\pm 4$)                & [18,20,21,22]\\
{\it HERMES\,J105751.1$+$573027 (HLSW$-$01)}                     &  2.9574    & 12.82                         & 13.17   & $10.9\pm 0.7$ ($9.2\pm 0.4$)                 & [18,23,24]\\
{\it SMM\,J12365$+$621226 (HDF\,850.1)}                          &  5.1830    & 12.43                         & 12.65	& 1.4                                          & [25]\\
{\it SMM\,J14009$+$0252}                                         &  2.9344    & 12.57                         & 12.74	& 1.5                                          & [14,26,27] \\
{\it SMM\,J140104.96$+$025223.5 (SMM\,J14011$+$0252)}            &  2.5653    & 12.19                         & 12.39	& $3.5\pm 0.5$ ($2.75\pm 0.25$)                & [28,29,30]\\
{\it H-ATLASJ142413.9$+$023040 (ID\.141)}                        &  4.2430    & 13.82                         & 14.09	& $4.6\pm 0.5$                                 & [31,32]\\
{\it SMM\,J163555.2$+$661150 (ABELL\,2218 Arc L)}                &  1.0313    & 11.16                         & 11.34	& 7.1                                          & [33]\\
{\it 1HERMESS350\,J170647.8$+$584623 (HFLS3)}                    &  6.3369    & 13.38                         & 13.72	& 1.0                                          & [33,34]\\ 
{\it SMMJ2135$-$0102 (Eyelash)}                                  &  2.3259    & 12.26                         & 12.36   & $32.5\pm 4.5$                                & [35,36]\\
{\it SPT-S233227$-$5358.5}                                       &  2.7256    & 12.77                         & 13.04   & $15\pm 5$                                    & [15]\\
\hline
                                                                 &            &                               &          & & \\
                                                                 &            &                               &          & & \\
\end{tabular}
    \begin{tablenotes}
      \small{
		\item[][1] \citet{bothwell2013a}; [2] \citet{alaghband-zadeh2013}; [3] \citet{greve2005}; [4] \citet{tacconi2006}; 
			   [5] \citet{frayer2008}; [6] \citet{engel2010}; [7] \citet{daddi2009}; [8] \citet{hodge2012}; [9] \citet{riechers2011a};
			   [10] \citet{ivison2011}; [11] \citet{neri2003}; [12] \citet{fu2013}; [13] \citet{ivison2010}; [14] \citet{thomson2012};
			   [15] \citet{aravena2013}; [16] \citet{bothwell2013b}; [17] \citet{ivison2013}; [18] \citet{bussmann2013};
			   [19] \citet{bussmann2013}; [20] \citet{lupu2012}; [21] \citet{harris2012}; [22] \citet{frayer2011}; [23] \citet{riechers2011b};
			   [24] \citet{conley2011}; [25] \citet{walter2012}; [26] \citet{weiss2009}; [27] \citet{harris2010}; [28] \citet{frayer1999}; 
			   [29] \citet{downes2003}; [30] \citet{sharon2013}; [31] \citet{cox2011}; [32] \citet{bussmann2012}; [33] \citet{riechers2013};
			   [34] \citet{robson2014}; [35] \citet{swinbank2010}; [36] \citet{danielson2011} 
}
    \end{tablenotes}
\label{table:high-z-sample}
\end{threeparttable}
\end{table*}

\section{Analysis}\label{section:analysis}
\subsection{SED fitting}
The pan-chromatic (FUV/optical to radio) spectral energy distributions (SEDs) of
our sample galaxies were modeled using CIGALE (Code Investigating GALaxy
Emission -- \citet{burgarella2005,noll2009}). CIGALE employs dust-attenuated
stellar population models to fit the FUV/optical SED, while at the same time
ensuring that the dust-absorbed UV photons are re-emitted in the FIR, thus maintaining 
energy-balance between the FUV and FIR. The FIR/sub-mm continuum is
modeled using the templates by \citet{dale2002} and \citet{chary2001}. For the
stellar emission population synthesis models from \citet{maraston2005} with a
Salpeter initial mass function were used, and for the reddening we used
attenuation curves from \citet{calzetti1994} with a wide range of V-band
attenuation values for young stellar populations. Despite having carefully
checked our samples against AGN, we allowed for the possibility of additional
dust emission from deeply buried AGN by including in our SED fits the 32 AGN
models from the \citet{fritz2006} library. Reassuringly, in no instances did the
AGN fraction exceeded $20\%$ of the total IR luminosity. Excellent fits were
obtained for all of the local galaxies due to their well-sampled SEDs. For the
high-$z$ galaxies, only sources with data points longward and shortward of (or
near) the dust peak ($\lambda_{\rm rest}\sim 100\,{\rm \mu m}$) were included in
the final analysis: a total of 35 out of the original 74 DSFGs. All SED
fits used in this paper can be found at \url{http://demogas.astro.noa.gr}, 
and will also be presented in a forthcoming paper (Xilouris et al., in prep.).

\begin{figure*}[h]
\includegraphics[scale=0.43]{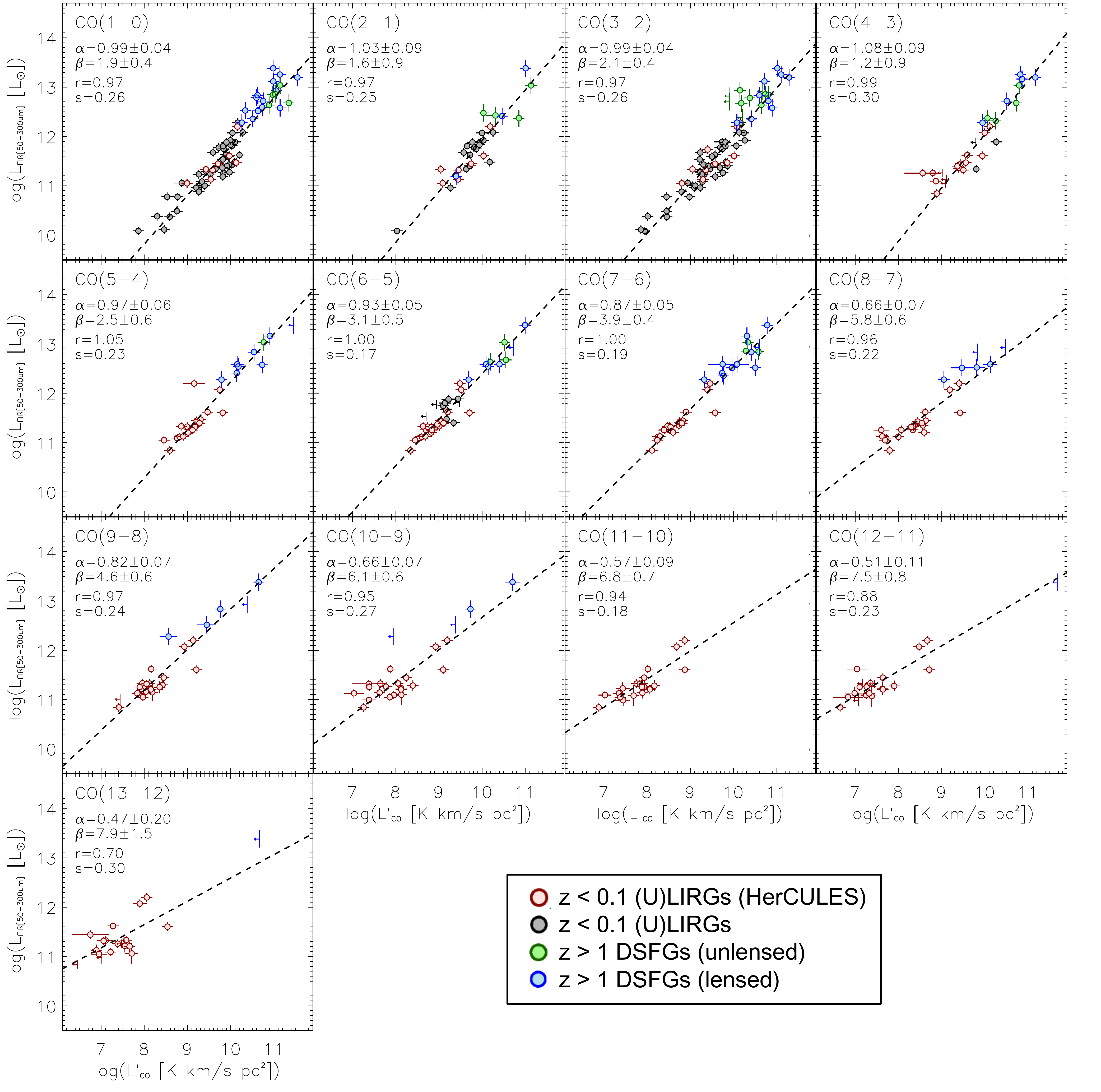}
\caption{$\log L_{\rm FIR}$ vs.\ $\log L'_{\rm CO}$ across the CO rotational
ladder (from $J=1-0$ to $J=13-12$). The low-$z$ ($z<0.1$) data include the
(U)LIRG sample from \citet{papadopoulos2012} (grey symbols) with CO observations
from $J=1-0$ to $J=6-5$, and (U)LIRGs from HerCULES (red symbols) observed in CO
$J=4-3$ to $J=13-12$ with {\it Herschel} SPIRE-FTS. Of the HerCULES sample, 11
sources have CO $J=1-0$ to $J=3-2$ coverage from \citet{papadopoulos2012} (see
\S~\ref{section:samples-and-data}).  The high-$z$ ($z>1$) sources are unlensed
DSFGs (green symbols) and strongly lensed ($\mu > 1$) DSFGs (blue symbols)
uncovered by various (sub)-mm surveys (\S~\ref{section:samples-and-data}).  The
dashed lines show the best fits of the function $\log L_{\rm FIR} = \alpha \log
L'_{\rm CO} + \beta$ to the data (\S~\ref{section:analysis}), with the optimum
parameter ($\alpha$, $\beta$) values and their errors indicated in each panel.
Also shown in each panel is the correlation coefficients ($r$) of the data and
their scatter ($s$) around the best-fit line.  The $L_{\rm FIR}$-values used
here were obtained by integrating the SEDs across the wavelength range
$50-300\,{\rm \mu m}$ (\S~\ref{section:analysis}) but near-identical relations
are obtained if instead the full IR-luminosity from $8-1000\,{\rm \mu m}$ is
used (Table \ref{table:fits}). Excluding the lensed DSFGs from the analysis did
not alter the best-fit values of $\alpha$ and $\beta$ significantly.}
\label{figure:fig1}
\end{figure*}
From the SED fits we derived the IR ($L_{\rm IR}$, from $8\,\rm{\mu m}$ to
$1000\,\rm{\mu m}$ rest-frame) and FIR ($L_{\rm FIR}$, from $50\,\rm{\mu m}$ to
$300\,\rm {\mu m}$ rest-frame) luminosities of our sample galaxies (Tables
\ref{table:low-z-sample} and \ref{table:high-z-sample}). We shall use the latter
for our analysis in order to minimize the effects of AGN, which are strongest in
the mid-IR regime (i.e., $\sim 8-40\,{\rm \mu m}$). Also, the mid-IR is rich in
PAH emission/absorption features, which could affect $L_{\rm IR}$ estimates. For
the uncertainty on our IR/FIR luminosity estimates we adopted the 1-$\sigma$
dispersion of the luminosity distributions obtained through bootstrapping
of the photometry errors 1000 times. Typical uncertainties, $\delta L_{\rm
FIR}$, were $\sim 20$ and $\sim 40$\% for the local and high-$z$ samples,
respectively, and were adopted across the board for the two samples. 
We stress that the above FIR luminosities are total luminosities, i.e.\ derived from aperture
fluxes that encompass the full extent of the galaxies, and thus match the
CO measurements.

\subsection{$L_{\rm FIR}-L'_{\rm CO}$ relations}
Fig.\ \ref{figure:fig1} shows the separate $\log L_{\rm FIR}-\log L'_{\rm CO}$
relations (where $\log$ is for base 10) for each CO transition (from CO $J=1-0$
to $J=13-12$) for the galaxy samples analyzed here. Highly significant correlations are
seen in all transitions, as given by their near unity linear correlation
coefficients ($r$, see Fig.\ \ref{figure:fig1}). Even for the highest transition
($J=13-12$), where the dynamical range spanned in luminosities is relatively
small, we see a statistically significant correlation. To ensure that the
observed correlations are not simply due to both $L_{\rm IR}$ and $L'_{\rm CO}$
being $\propto D_{\rm L}^2$ (the luminosity distance squared), we calculated for
each correlation the partial Kendall $\tau$-statistic \citep{akritas1996} with
$D_{\rm L}^2$ as the test variable. In all cases (up to $J=13-12$), we find
probabilities $P < 10^{-6}$ that the observed FIR$-$CO correlations are falsely
induced by the fact that $luminosity \propto D_L^2$.
%
\begin{table}
\centering
\caption{Best-fit slopes ($\alpha$) and intersection points ($\beta$), along
with the associated scatter of the data around the best-fit relation, inferred
from Fig.\ \ref{figure:fig1}. The corresponding values using $L_{\rm
IR[8-1000\,\mu m]}$ instead of $L_{\rm FIR[50-300\mu m]}$ are given in
parentheses.}
\begin{tabular}{lccc}
\hline
Transition          & $\alpha$         & $\beta$        &  $s$       \\
                    &                  &                &            \\
\hline
\hline
CO$(1-0)$			& $0.99\pm 0.04$   & $1.9\pm 0.4$   & $0.26$     \\	
        			& $(1.00\pm 0.05)$ & $(2.0\pm 0.5)$ & $(0.27)$   \\	

CO$(2-1)$			& $1.03\pm 0.09$   & $1.6\pm 0.9$   & $0.25$     \\	
        			& $(1.05\pm 0.10)$ & $(1.7\pm 0.9)$ & $(0.27)$   \\	

CO$(3-2)$			& $0.99\pm 0.04$   & $2.1\pm 0.4$   & $0.26$     \\	
        			& $(1.00\pm 0.05)$ & $(2.2\pm 0.5)$ & $(0.28)$   \\	

CO$(4-3)$			& $1.08\pm 0.09$   & $1.2\pm 0.9$   & $0.30$     \\	
        			& $(1.08\pm 0.09)$ & $(1.5\pm 0.9)$ & $(0.29)$   \\	

CO$(5-4)$			& $0.97\pm 0.06$   & $2.5\pm 0.6$   & $0.23$     \\	
        			& $(0.97\pm 0.06)$ & $(2.8\pm 0.6)$ & $(0.23)$   \\	

CO$(6-5)$			& $0.93\pm 0.05$   & $3.1\pm 0.5$   & $0.17$     \\	
        			& $(0.95\pm 0.06)$ & $(3.2\pm 0.5)$ & $(0.18)$   \\	

CO$(7-6)$			& $0.87\pm 0.05$   & $3.9\pm 0.4$   & $0.19$     \\	
        			& $(0.87\pm 0.05)$ & $(4.1\pm 0.4)$ & $(0.19)$   \\	

CO$(8-7)$			& $0.66\pm 0.07$   & $5.8\pm 0.6$   & $0.22$     \\	
        			& $(0.66\pm 0.07)$ & $(6.1\pm 0.6)$ & $(0.20)$   \\	

CO$(9-8)$			& $0.82\pm 0.07$   & $4.6\pm 0.6$   & $0.24$     \\	
        			& $(0.85\pm 0.07)$ & $(4.6\pm 0.6)$ & $(0.22)$   \\	

CO$(10-9)$			& $0.66\pm 0.07$   & $6.1\pm 0.6$   & $0.27$     \\	
        			& $(0.69\pm 0.08)$ & $(6.1\pm 0.6)$ & $(0.27)$   \\	

CO$(11-10)$			& $0.57\pm 0.09$   & $6.8\pm 0.7$   & $0.18$     \\	
        			& $(0.61\pm 0.09)$ & $(6.8\pm 0.7)$ & $(0.17)$   \\	

CO$(12-11)$			& $0.51\pm 0.11$   & $7.5\pm 0.8$   & $0.23$     \\	
        			& $(0.55\pm 0.11)$ & $(7.5\pm 0.8)$ & $(0.23)$   \\	

CO$(13-12)$			& $0.47\pm 0.20$   & $7.9\pm 1.5$   & $0.30$     \\	
        			& $(0.51\pm 0.21)$ & $(7.9\pm 1.6)$ & $(0.31)$   \\	
\hline
                    &                  &                &            \\
                    &                  &                &            \\
\end{tabular}
\label{table:fits}
\end{table}

A function of the form $\log L_{\rm FIR} = \alpha \log L'_{\rm CO} + \beta$ was
adopted to model the correlations, and the optimal values of the model
parameters ($\alpha$ and $\beta$) were fitted (to this end we used the IDL routine {\tt linmix\_err},
\citet{kelly2007}). The slopes ($\alpha$) and intersection points ($\beta$)
inferred from fits to the combined low- and high-$z$ samples are given in Table
\ref{table:fits}, along with the scatter ($s$) of the data-points around the
fitted relations. The fits are shown as dashed lines in Fig.\ \ref{figure:fig1}.
The best-fit $(\alpha,\beta)$-values obtained by using $L_{\rm IR}$ instead of
$L_{\rm FIR}$ are also listed in Table \ref{table:fits}. Within the errors, the
fitted parameters are seen to be robust against the adopted choice of FIR or IR
luminosity. Also, our results did not change in any significant way when
omitting the lensed DSFGs from the analysis. Often lensing amplification factors
are uncertain, and strong lensing can not only skew the selection of sources
towards more compact (and thus more likely warm) sources, but for a given
source it may also boost the high-$J$ CO lines relative to to the lower lines
(this is discussed further in \S~\ref{subsection:caveats}).

Figs.\ \ref{figure:fig2} and \ref{figure:fig3} show the slopes and
normalisations, respectively, of the $\log L_{\rm FIR}-\log L'_{\rm CO}$
relations derived above as a function of the critical densities probed by the
various CO transitions. The critical densities are calculated as $n_{\rm crit} =
A_{ul}/\sum_{i\ne u} C_{ui}$, where $A_{\rm ul}$ is the Einstein coefficient for
spontaneous decay, and $\sum_{i\ne u} C_{ui}$ is the sum over all collisional
coefficients (with H$_2$ as the collisional partner) out of the level $u$,
`upwards' and `downwards' (see Table \ref{table:critical-densities} where, as a
reference, we also list $n_{\rm crit}$-values for a number of HCN and CS
transitions). Although, it is the first three levels `up' or `down' from the
$u$-level (i.e., $| u - i | < 3$) that dominate the sum, often in the literature
molecular line critical densities are calculated for a two-level system only
(i.e., $|u - i| = 1$), or for the downward transitions only -- both practices
that can significantly overestimate the true $n_{\rm crit}$ for a given
transition. The collision rates were adopted from the Leiden Atomic and
Molecular Database (LAMDA; \citet{schoeier2005}) for $T_{\rm k}=40\,{\rm K}$,
which is within the range of typical dust and gas temperatures encountered in
local (U)LIRGs and high-$z$ DSFGs (e.g., \citet{kovacs2006}).  We do not correct
for optical depth effects (i.e., line-trapping) as these are subject to the
prevailing average ISM conditions, but we note that large optical depths
(especially for low-$J$ CO and HCN lines) can significantly lower the effective
critical density to: $n^{(\beta)} _{\rm crit}=\beta_{ul} n_{\rm crit}$ where
$\beta_{ul}$ is the average line escape probability
($=[1-\exp(-\tau_{ul})]/\tau_{ul}$ for spherical geometries). Of course, the
collisional excitation of CO to higher rotational states ($E_{\rm J}$) is set
not only by the gas density but also by its kinetic temperature.  The minimum
temperature ($T_{\rm min}$) required for significant collisional excitation of a
given rotational state is approximately given by: $\sim E_{\rm J}/k_{\rm B} =
B_{\rm rot} J (J+1)/k_{\rm B}$, where $B_{\rm rot}$ is the rotational constant
of CO, and $k_{\rm B}$ is the Boltzmann constant.  As a rule of thumb, high
kinetic temperatures are needed in order to excite the high-$J$ CO lines (see
Table \ref{table:critical-densities}), although due to the $n-T_{\rm k}$
degeneracy this can also be achieved for very dense, low-temperature gas.
%
\begin{table}
\centering
\caption{Critical densities ($n_{\rm crit}$) and upper level energies
($E_{J}/k_{\rm B}$) of the rotational ladder of CO, and selected transitions of
HCN and CS, assuming H$_2$ is the main collision partner. The $n_{\rm
crit}$-values are calculated for a kinetic temperature of $T_{\rm k}=40\,{\rm
K}$, and an ortho-H$_2$ : para-H$_2$ ratio of 3.}
\begin{tabular}{lcc}
\hline
Transition          & $n_{\rm crit}$      & $E_{J}/k_{\rm B}$   \\
                    & $\rm [cm^{-3}]$     & $\rm [K]$      \\
\hline
\hline
CO($1-0$)			& $3.09\times10^{2}$  & $5.53$         \\	
CO($2-1$)			& $2.73\times10^{3}$  & $16.60$        \\	
CO($3-2$)			& $9.51\times10^{3}$  & $33.19$        \\	
CO($4-3$)			& $2.29\times10^{4}$  & $55.32$        \\	
CO($5-4$)			& $4.48\times10^{4}$  & $82.97$        \\	
CO($6-5$)			& $7.70\times10^{4}$  & $116.16$       \\	
CO($7-6$)			& $1.21\times10^{5}$  & $154.87$       \\	
CO($8-7$)			& $1.78\times10^{5}$  & $199.11$       \\	
CO($9-8$)			& $2.50\times10^{5}$  & $248.88$       \\	
CO($10-9$)			& $3.41\times10^{5}$  & $304.16$       \\	
CO($11-10$)			& $4.63\times10^{5}$  & $364.97$       \\	
CO($12-11$)			& $6.00\times10^{5}$  & $431.29$       \\	
CO($13-12$)			& $7.55\times10^{5}$  & $503.13$       \\	
\hline
HCN($1-0$)			& $1.07\times10^{5}$  & $4.25$         \\	
HCN($2-1$)			& $1.02\times10^{6}$  & $12.76$        \\	
HCN($3-2$)			& $3.52\times10^{6}$  & $25.52$        \\	
HCN($4-3$)			& $8.84\times10^{6}$  & $42.53$        \\	
\hline
CS($1-0$)			& $6.77\times10^{3}$  & $2.35$         \\	
CS($2-1$)			& $6.50\times10^{4}$  & $7.05$         \\	
CS($3-2$)			& $2.40\times10^{5}$  & $14.11$        \\	
CS($5-4$)			& $1.34\times10^{6}$  & $35.27$        \\	
CS($6-5$)			& $2.36\times10^{6}$  & $49.37$        \\	
CS($7-6$)			& $3.76\times10^{6}$  & $65.83$        \\	
\hline
                    &                  &                   \\
                    &                  &                   \\
\end{tabular}
\label{table:critical-densities}
\end{table}

Two trends regarding the $L_{\rm FIR}-L'_{\rm CO}$ relations become apparent
from Figs.\ \ref{figure:fig2} and \ref{figure:fig3} (see also Table
\ref{table:fits}). Firstly, the slopes are linear for $J=1-0$ to $J=5-4$ but
then start becoming increasingly sub-linear, the higher the $J$ level.
Secondly, the normalization parameter $\beta$ remains roughly constant ($\sim
2$) up to $J=4-3$, $5-4$, but then increases with higher $J$ level, reaching
$\beta \sim 8$ for $J=13-12$, which for a given CO luminosity translates into
$\sim 6$ orders of magnitude higher $L_{\rm FIR}$. We stress that although the
$L_{\rm FIR}-L'_{\rm CO}$ relations are linear, and $\beta$ roughly constant, up
to $J=5-4$, it does not in general imply that the CO lines are thermalized
(i.e., $L'_{\rm CO_{\rm J,J-1}}/L'_{\rm CO_{1,0}}\simeq 1$) up to this
transition.  There is significant scatter within the samples, and while a few
sources do have nearly-thermalized $J=2-1$, $3-2$, and/or $4-3$ lines, in
general, $L'_{\rm CO_{\rm J,J-1}}/L'_{\rm CO_{1,0}} \ls 1$. In fact, re-writing
the $L_{\rm FIR}-L'_{\rm CO}$ relations as:
\begin{equation}
L'_{\rm CO_{\rm J,J-1}}/L'_{\rm CO_{\rm 1,0}} = L_{\rm FIR}^{\alpha_{\rm J,J-1}^{-1} - \alpha_{\rm 1,0}^{-1}} \times
10^{\beta_{\rm 1,0} - \beta_{\rm J,J-1}},
\end{equation}
and inserting the fitted values from Table \ref{table:fits} yields $L'_{\rm
CO_{\rm J,J-1}}/L'_{\rm CO_{\rm 1,0}} < 1$ over the range $L_{\rm FIR} =
10^{9-14}\,\Lsolar$.

In the following sections we discuss these empirical relations in the context of
existing theoretical models and the new observational studies of such relations
using heavy rotor molecules like HCN, and CS.

\section{The slope of the $L_{\rm FIR}-L'_{\rm CO}$ relations}\label{section:alpha}
\subsection{Comparison with previous studies}\label{subsection:comparison-of-slopes}
Before comparing our derived FIR$-$CO slopes with those from the literature we
must add two cautionary notes, namely: a) many studies examine the $L'_{\rm
CO}-L_{\rm FIR}$ relation, rather than $L_{\rm FIR}-L'_{\rm CO}$, and one
cannot compare the two simply by inferring the inverse relation, b) often, only
the errors in one variable (typically $L'_{\rm CO}$) are taken into account when
fitting such relations, when in fact the uncertainties in both $L_{\rm FIR}$ and
$L'_{\rm CO}$ must be considered (see \citet{mao2010} for a further discussion).
Failing to do so can result in erroneous estimates of the slope.

For these reasons, we have re-fitted the data from a number of studies (see
below) using the method described in \S~\ref{section:analysis}, i.e., with errors
in both $L_{\rm FIR}$ and $L'_{\rm CO}$ and including only sources with $L_{\rm
FIR} \gs 10^{11}\,\Lsolar$ in our analysis. Finally, not all studies use the
FIR definition used here to infer $L_{\rm FIR}$, while other studies use the
full ($8-1000\,{\rm \mu m}$) luminosity. These differences can result in a
different overall normalization (i.e., $\beta$), but are not expected to affect
the determination of $\alpha$ (see Table \ref{table:fits} where there is little
change in $\alpha$ when switching between $L_{\rm FIR[50-300\mu m]}$ and $L_{\rm
IR[8-1000\mu m]}$).

\begin{figure*}[t]
\begin{center}
\includegraphics[width=0.75 \textwidth]{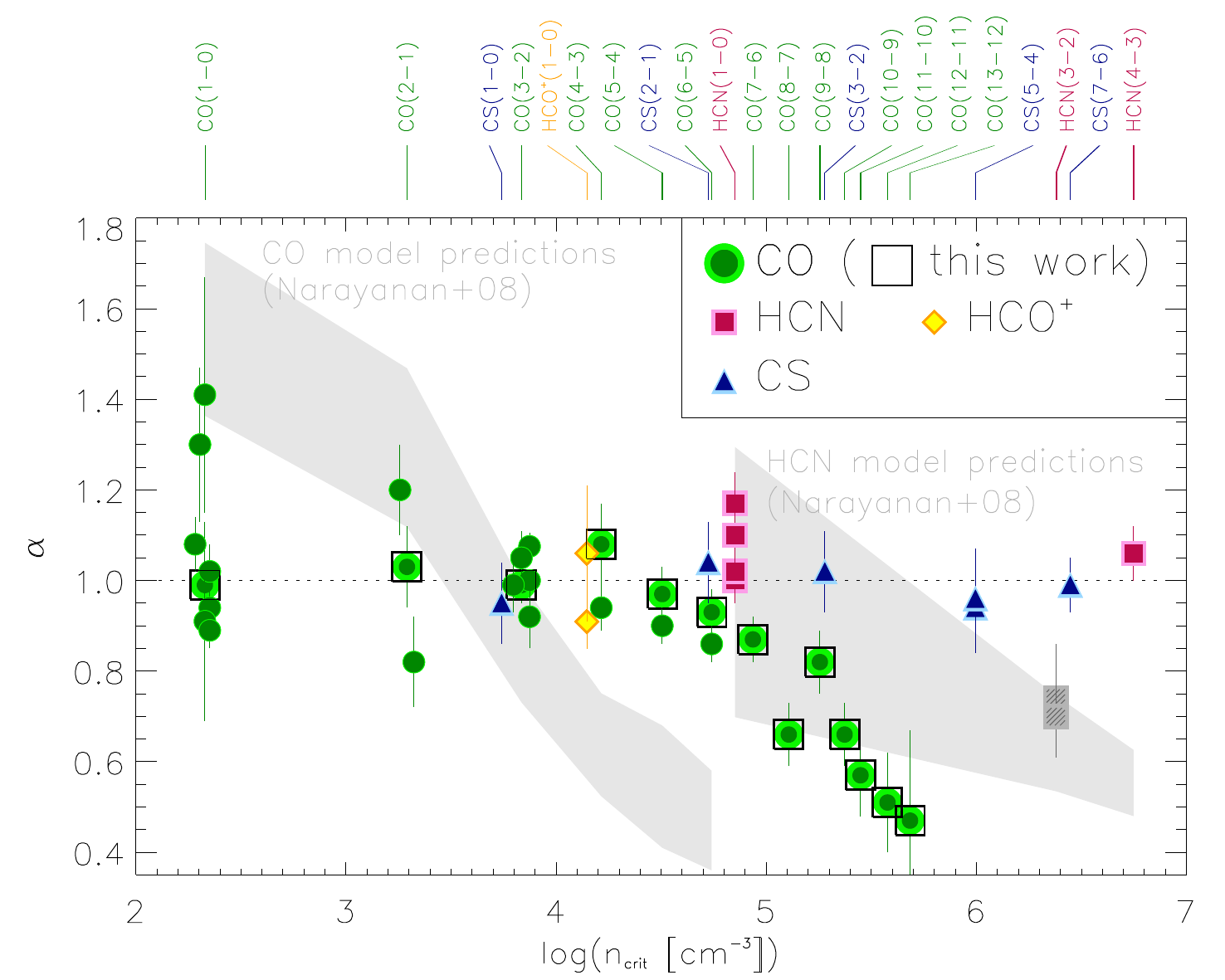}
\end{center}
\caption{A compilation of slope ($\alpha$) determinations from the
literature for CO
\citep{yao2003,gao2004b,narayanan2005,baan2008,juneau2009,iono2009,bayet2009,genzel2010,mao2010},
HCO$^+$ \citep{baan2008,garcia-burillo2012}, HCN
\citep{gao2004b,wu2005,wu2010,bussmann2008,GC-B-2008,juneau2009,garcia-burillo2012,zhang2014},
and CS \citep{wu2010,wang2011,zhang2014} (the CS $J=1-0$, $2-1$, $3-2$, and
$5-4$ results are from Zhang et al., in prep.). In some cases
\citep{yao2003,baan2008} we had to re-fit the FIR$-$CO relations as to
facilitate a direct comparison with our findings (see \S~\ref{subsection:comparison-of-slopes}).
The slopes derived in this paper are outlined by black squares. For the first
three CO transitions, the different $\alpha$-estimates have been slightly offset
horizontally in order to ease the comparison. The grey-shaded regions show the
CO (left) and HCN(right) slopes (within a 1-$\sigma$ scatter) predicted by one
of two theoretical models \citep{narayanan2008}.  The $L_{\rm FIR}-L'_{\rm X}$
relations for lines with high $n_{\rm crit}$ but low $E_{J}/k_{\rm B}$ (X=HCN,
CS) have slopes consistent with unity across the critical density regime of
$\sim 10^{4-7}\,{\rm cm^{-3}}$, and are inconsistent with the theoretical
predictions. A statistically significant trend of $\alpha$ versus $n_{\rm crit}$
is found for the $L_{\rm IR}-L'_{\rm CO}$ relation with $\alpha \simeq 1$ up to CO
$J=5-4$ but then decreasing with higher $J$ (and thus $n_{\rm crit}$). However
the high $E_{J}/k_{\rm B}$ values of the CO $J=6-5$ to $J=13-12$ transitions
($\sim 115-500\,{\rm K}$) place them well outside the applicability of both
current theoretical models and thus this trend cannot be used to test them.
Following the reasoning laid out by \citet{krumholz2007}, the sub-linear slopes
of the $L_{\rm FIR}-L'_{\rm CO}$ relations for such high-$J$ CO lines are
actually quite unexpected (see discussion). We ignore the FIR$-$HCN($3-2$)
slopes inferred by \citet{bussmann2008} and \citet{juneau2009} (shown as
grey-hatched squares) since their data were not appropriately beam-corrected
(see \S~\ref{subsection:alpha-ncrit}).}
\label{figure:fig2}
\end{figure*}

From Fig.\ \ref{figure:fig2} we note the overall good agreement between the
FIR$-$CO slopes derived here and values from the literature. For CO($1-0$),
however, one set of measurements found super-linear slopes ($\alpha_{{\rm
CO}_{1,0}} \sim 1.3-1.4$; \citet{juneau2009,bayet2009}), while most others favor
a slope of unity \citep[this work]{gao2004b,mao2010,ivison2011}.  Note, our
re-analysis of the \citet{yao2003} and \citet{baan2008} data revised their
slopes from super-linear to linear: $\alpha_{{\rm CO}_{1,0}} = 0.94\pm 0.07$ and
$1.08\pm 0.06$, respectively (a similar result was found by \citet{mao2010}).
\cite{gao2004b} finds a super-linear FIR$-$CO slope ($\alpha = 1.3-1.4$) from
their entire sample (combining $L_{\rm IR} \sim 10^{10}\,\Lsolar$ objects with
LIRGs and ULIRGs), yet including only the LIRGs and ULIRGs in the analysis, we
obtain a linear slope ($\alpha_{{\rm CO}_{1,0}} = 0.91\pm 0.22$). Re-fitting the
data presented in \citet{juneau2009} and \citet{bayet2009} we reproduce their
super-linear slopes.

For CO($2-1$) our slope of unity is consistent within the errors with
\citet{bayet2009}, who finds a slightly super-linear slope based on 17 sources.
Our re-analysis of the CO$(2-1)$ data by \citet{baan2008} yields a sub-linear
slope of $\alpha=0.82\pm 0.10$. However, as pointed out by the authors
themselves, a non-negligible fraction of the total CO$(2-1)$ emission is likely
to have been missed due to the smaller telescope beam at higher frequencies --
thus biasing the FIR$-$CO relation to a shallower value of $\alpha$. 

In the case of CO($3-2$), the existing slope-determinations
\citep{yao2003,narayanan2005,iono2009,bayet2009,mao2010}, including our own, are
all in agreement and favor a value of unity within the errors. This includes a
re-analysis of the \citet{yao2003} data, which yielded $\alpha = 1.00$ (see also
\citet{mao2010}).

For CO($4-3$) to CO($7-6$) there is agreement within the errors between our
results and the slopes found by \citet{bayet2009}, which however were determined
using observed as well as model-extrapolated CO luminosities of 7 low-$z$ and 10
high-$z$ sources. A departure from linear towards sub-linear is also found by
the latter study, albeit we find this turn-over to occur at $J=6-5$
rather than at $J=4-3$ deduced by \citet{bayet2009}. Here we must note, however,
that the use of models to extrapolate to high-$J$ CO luminosities is not safe
and artificial turnovers can be introduced because of the inability of such
models -- in the absence of appropriate line data -- to reliably account for the
existence of warmer and denser gas components. This further underscores the
value of our observed $L_{\rm FIR}-L'_{\rm CO}$ relations from $J=1-0$ to
$J=13-12$ in safely determining such departures from linearity and/or in
normalization before proceeding towards any interpretation based on ISM/SF
physics.
 
\subsection{Super-linear slopes: the simplest scenario }\label{subsection:S-K simple}
The few super-linear slopes of FIR$-$CO luminosity relations for low-$J$ CO
lines that survive careful re-analysis \citep{juneau2009,bayet2009} could be a
byproduct of dense molecular gas being the direct SF fuel in all galaxies (with
a constant SFE) and a $f_{\rm dense,X}=M_{\rm dense}/M_{\rm X}$ (where X could
be total H$_2$ gas mass traced by CO $J=1-0$, $J=2-1$ lines)
that varies within the galaxy sample with $d(f_{\rm dense})/dL_{\rm FIR}>0$.
Variations of this simple scenario have been suggested throughout the literature
\citep{wong2002,gao2004b}, and unlike more sophisticated interpretations of such
super-linear slopes offered by the two theoretical treatises available on this
matter \citep{krumholz2007,narayanan2008}, the only assumption here is that
$d(f_{\rm dense})/dL_{\rm FIR}>0$. The latter is a well-documented fact as
starbursts/mergers, which dominate the high-$L_{\rm FIR}$ end, are observed to
have larger dense/total gas mass fractions than lower-$L_{\rm FIR}$ isolated
disks (e.g., \citet{gao2004b,garcia-burillo2012}).

Within this scheme, the further the gas phase X is from the dense, star forming
phase in terms of physical conditions and relevance to the star formation, the
higher the $d(f_{\rm dense})/dL_{\rm FIR}$ value is and the deduced super-linear
slope of the $L_{\rm FIR}-L'_{\rm X}$ (or corresponding S-K) relation.  On the
other hand for galaxy samples with a smaller range of IR-luminosities that have
a nearly constant $f_{\rm dense,X}$  -- such as the (U)LIRG+DSFG sample
considered in this paper (see \S~\ref{subsection:beta}) -- a linear slope of an
$L_{\rm FIR}-L'_{\rm X}$ relation can still be recovered for e.g., CO $J=1-0$
that traces all metal-rich molecular gas mass rather than only the dense SF one.
We return to this point later in our discussion.

\begin{figure*}[t]
\begin{center}
\includegraphics[width=0.75 \textwidth]{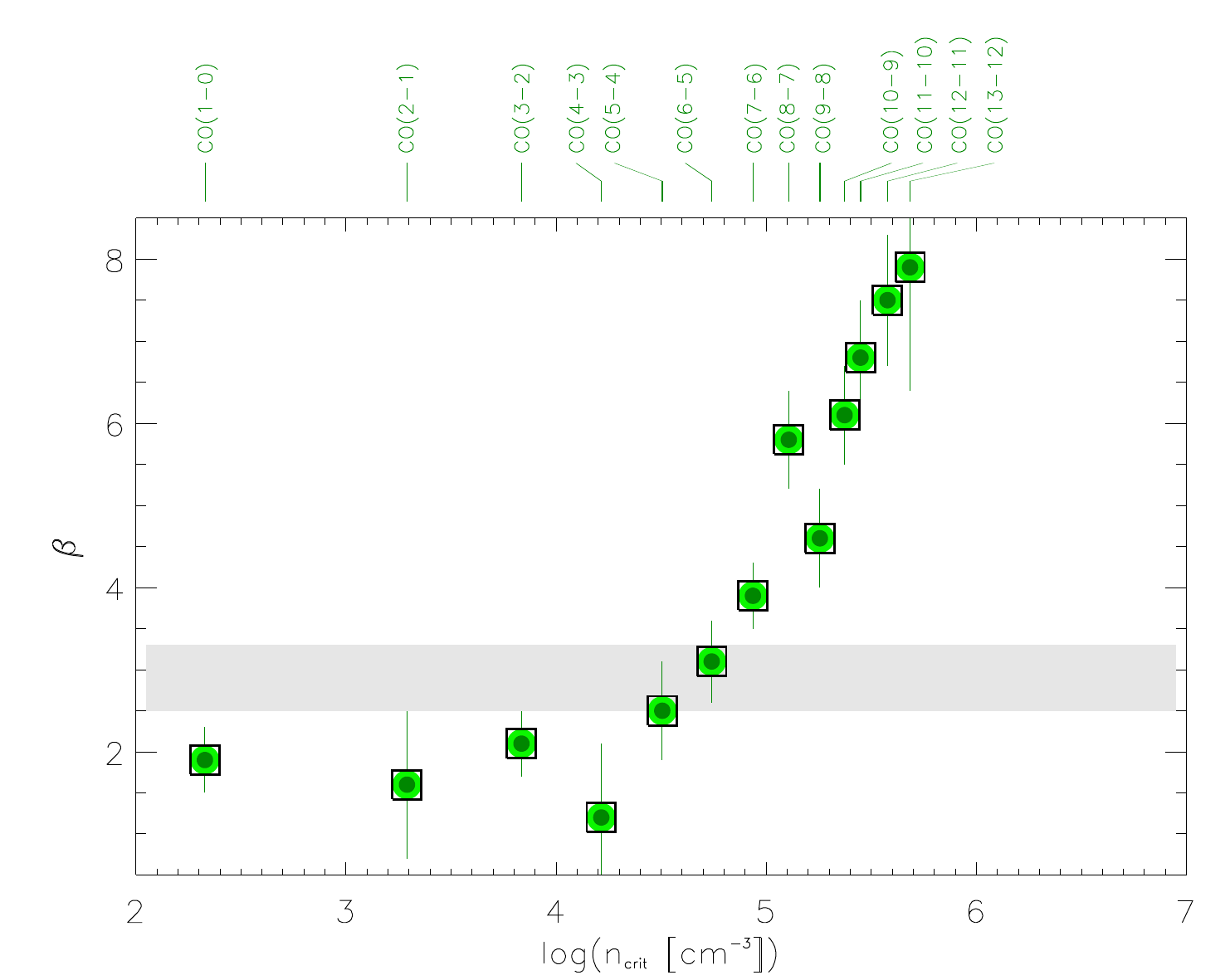}
\end{center}
\caption{The normalisations ($\beta$) of the FIR$-$CO relations presented
in this paper (Fig.\ \ref{figure:fig1} and Table \ref{table:fits}). For $J=1-0$
to $5-4$ the normalisations are constant (within the errors), with values
($\beta\sim 1.2-2.5$) close to that expected from Eddington limited star
formation ($\sim 2.5-3.3$, shown as grey shaded area), assuming a Rosseland-mean
opacity in the range $5-30\,{\rm cm^2\,g^{-1}}$ and a CO-to-H$_2$ conversion
factor of $0.8\,{\rm K\,km\,s^{-1}\,pc^2}$ (see also \S~\ref{subsection:beta}).
For $J=6-5$ and higher, the $\beta$-values increase with $J$ as the FIR$-$CO
slopes become sub-linear.
}
\label{figure:fig3}
\end{figure*}

\section{Confronting theoretical models}
\subsection{$\alpha$ versus $\lowercase{n_{\rm crit}}$}\label{subsection:alpha-ncrit}
As already mentioned in the introduction, the S-K relation, and especially the
important one involving the dense gas component in galaxies, is not easily
accessible observationally. So one falls back to the much more observationally
accessible proxy $L_{\rm FIR}-L'_{\rm X}$ relations. Their index can then be
linked to that of an assumed underlying S-K relation in galaxies using two
theoretical models \citep{krumholz2007,narayanan2008}. Both models posit the
same intrinsic S-K relation of $\rho _{\rm SFR}\propto \rho^{1.5}_{\rm gas}$ (or
$\Sigma_{\rm SFR}\propto \Sigma^{1.5}_{\rm gas}$ (for disks of near-constant
scale-height), justified under the assumption of a constant gas fraction
transformed into stars per free fall time ($t_{\rm ff}\propto (G\rho_{\rm
gas})^{-1/2}$). The same S-K relation emerges also if the SF timescale is
instead set by the dynamical timescale of a marginally Toomre-stable galactic
disk with SF converting a fixed fraction of gas into stars over such a timescale
\citep{elmegreen2002}. 

Both models give expected values of $\alpha$ versus $n_{\rm crit}({\rm X})$ for
$L_{\rm FIR}-L'_{\rm X}$ luminosity relations over a large range of line
critical densities, but both models are applicable only for lines that require
low temperatures to excite $E_{J}/k_{\rm B}\la 30\,{\rm K}$ (for the
\citet{krumholz2007} model this limit is $\sim 10\,{\rm K}$). The reasons behind
this limitation are explicit assumptions about isothermal gas states at a set
temperature \citep{krumholz2007}, or the tracking of such states over a small
range of $T_{\rm k}\sim 10-30\,{\rm K}$ \citep{narayanan2008}. The low-$J$ lines
of heavy rotor molecular line data-sets found in the literature and our low-$J$
CO lines are certainly within the range of applicability of these models. The
indices of the corresponding $L_{\rm FIR}-L'_{\rm X}$ power law relations can
thus be compared to these theoretical predictions.

\citet{gao2004a,gao2004b} found a linear FIR$-$HCN correlation for the
$J=1-0$ transition (see also \citet{baan2008}), which extends from local
ULIRGs/LIRGs ($L_{\rm IR} \sim 10^{11-12}\,{\rm \Lsolar}$) to normal, star
forming galaxies ($L_{\rm IR} \sim 10^{9-10}\,{\rm \Lsolar}$), down to
individual Galactic molecular clouds with $L_{\rm IR} \gs 10^{4.5}\,{\rm
\Lsolar}$ \citep{wu2005,wu2010}. A weakly super-linear FIR$-$HCN($1-0$) slope
($\alpha \simeq 1.2$) was found by \citet{GC-A-2008} and
\citet{garcia-burillo2012} over a combined sample of local normal galaxies and
LIRG/ULIRGs. A careful analysis by \citet{garcia-burillo2012}, however,
demonstrated that a bimodal fit (i.e., a different normalization parameter
$\beta $) is better, with each galaxy sample well fit by a linear relation.
Finally, a weakly super-linear $L_{\rm FIR}-L_{\rm HCN_{1,0}}$ appears when
high-$z$ observations of the most IR-luminous starburst galaxies and QSOs are
included in the locally-established relation \citep{gao2007,riechers2007}. This
could also be bimodal instead, but with an otherwise linear $L_{\rm FIR}-L_{\rm
HCN_{1,0}}$ relation (and with insufficient high-$L_{\rm FIR}$ objects to decide
the issue).  A physical reason for such bimodalities is discussed in
\S~\ref{subsection:beta}.

Extending HCN observations to include many more objects in the crucial $L_{\rm
IR}>10^{12}\,{\rm \Lsolar}$ regime is necessary for deciding such issues. Even
then one must eventually obtain the underlying SFR$-M_{\rm dense}$ relation
before arriving at secure conclusions about a varying ${\rm SFE} = {\rm
SFR}/M_{\rm dense}$ of the dense gas in (U)LIRGs. The latter is the crucial
physical quantity underlying the normalization of such $L_{\rm FIR}-L'_{\rm
HCN}$ relations and e.g., a rising or bimodal $X_{\rm HCN}=M_{\rm dense}/L_{\rm
HCN_{1,0}}$ factor towards high-$L_{\rm IR}$ systems can easily erase purported
SFE trends obtained by using single-line proxies of dense gas.

A sub-linear FIR$-$HCN slope ($\sim 0.7-0.8$) for the $J=3-2$ transition has
been reported \citep{bussmann2008,juneau2009} but is very likely biased low due
to not having performed any beam correction (see
\S~\ref{section:samples-and-data}) for some of their very nearby extended
objects, where the HCN beam does not cover the entire IR emitting region.  For
these sources, the HCN measurements do not match the IR luminosities, and since
they all reside at the lower end of the HCN luminosity distribution, the net
effect will be to bias the relation towards shallower values.  For this reason
we have chosen to ignore the sub-linear FIR$-$HCN slopes from
\citet{bussmann2008} and \citet{juneau2009}.  A recent survey of HCN $J=4-3$ and
CS $J=7-6$ \citep{zhang2014}, and CS $J=1-0$, $2-1$, $3-2$, and $5-4$ (Zhang et
al., in prep.), towards nearby star forming galaxies ($L_{\rm IR}\sim
10^9-10^{12}\,{\rm \Lsolar}$), where such effects have been adequately accounted
for, establishes a slope $\alpha \sim 1$ for these transitions (see also
\citet{wu2010} and \citet{wang2011}). Many of these CS transitions have higher
critical densities than HCN, and CS is furthermore less prone to IR pumping
effects than HCN is (CS is pumped at $7.9\,{\rm \mu m}$ compared to $14\,{\rm
\mu m}$ for HCN). Pumping of HCN (and also HNC), however, typically only becomes
important at dust temperatures $\gs 50\,{\rm K}$ \citep{aalto2007}, and would
typically require even higher temperatures for CS.  This is important since
pumping could affect the CS/HCN luminosities (especially at high-$J$), and thus
in principle result in linear IR-CS/HCN relations. In Fig.\ \ref{figure:fig2} we
summarize all the observationally determined $L_{\rm IR}-L'_{\rm HCN}$ and
$L_{\rm IR}-L'_{\rm CS}$ slopes from the literature along with those derived
from our CO lines.  Overall, the data suggest $\alpha \sim 1$ for our $L_{\rm
FIR}-L'_{\rm CO}$ relations from $J=1-0$ up to $J=5-4$, $6-5$ and for the
heavy-rotor molecular lines. The latter cover a range of $n_{\rm crit}\sim
10^{4}-10^{7}\,{\rm cm^{-3}}$, i.e., reaching up well into the high-density
regime of the star forming gas phase.

\bigskip

From Fig.\ \ref{figure:fig2} it becomes clear that most observations are
incompatible with current model predictions (shown as the grey-shaded area) both
for the heavy rotor and the low-$J$ CO lines (where such models remain
applicable). Super-linear slopes do appear for some CO $J=1-0$ data-sets
but then, unlike model predictions, the slopes remain linear for lines with much
higher critical densities, including those of mid-$J$ CO lines $J=3-2, 4-3, 5-4$
(the FIR$-$CO luminosity relation for $J=6-5$ is also compatible with a linear
one).  

In summary, we conclude that the global $L_{\rm FIR}$-$L'_{\rm X}$ relations
in (U)LIRGs and DSFGs, as parametrized by $\log L_{\rm FIR} = \alpha \log
L'_{\rm X} + \beta$, are linear for heavy rotor lines, and our CO line
data-set up to $J=6-5$, at which point the relations become increasingly
sub-linear for higher $J$. Moreover, the normalization factor $\beta $ shows a
similar behavior by being nearly constant $\beta \sim 2$ up to $J=5-4/6-5$ but
then starting to increase systematically with increasing $J$, reaching $\beta
\sim 8$ for $J=13-12$.

\subsection{More ISM physics in $\beta$ rather than in $\alpha$?}\label{subsection:beta}
The simplest scenario outlined in \S~\ref{subsection:S-K simple} seems to work
both for the low-$J$ CO and the heavy rotor molecular lines with much higher
critical densities. Given that our sample solely consists of (U)LIRGs (i.e.,
$L_{\rm IR[8-1000\,\mu m]}\ge 10^{11}\,\Lsolar$) for which the dense gas
fraction (i.e., $f_{\rm dense} = L'_{\rm HCN_{\rm low-J}}/L'_{\rm CO_{\rm
low-J}}$) is nearly constant (see discussion in
\S~\ref{section:alpha-and-beta}), linear slopes are to be expected for its
low-$J$ FIR$-$CO relations. For other samples in the literature (e.g.,
\citet{bayet2009}) that reach lower IR luminosities ($\sim 10^{10}\,\Lsolar$)
and thus span a wider range in $L_{\rm IR}$ (over which $f_{\rm dense}$ changes
appreciably) the super-linear slopes of their low-$J$ FIR$-$CO relations seen in
Fig.\ \ref{figure:fig2} are also expected. In this simple picture neither the
occasional super-linear nor the linear slope of the $L_{\rm FIR}-L'_{\rm CO}$
low-$J$ relations carry any profound ISM physics other than more dense gas mass
corresponds to proportionally higher SFRs (a picture also suggested by
\citet{gao2004b} and \citet{wu2005}). 

For dense gas tracer lines this shows itself directly with $L_{\rm IR}-L'_{\rm
X}$ relations that always have linear slopes.  In this picture there is
actually more ISM physics to be found in exploring what sets the value of the
normalization parameter $\beta$ rather than the slope of $L_{\rm IR}-L'_{\rm
X}$ relations.  

The low- and high-$z$ (U)LIRGs studied here are highly dust-obscured
galaxies, and radiation pressure exerted by the strong absorption and scattering
of FUV light by dust grains could be an important feedback mechanism, possibly
setting the value of the normalization (and ultimately regulating the SF). The maximum attainable
$L_{\rm IR}/M_{\rm dense}$ ratio of a star forming region before radiation
pressure halts higher accretion rates is ultimately set by the Eddington limit
giving: $L_{\rm IR}/M_{\rm dense}\sim 500\,{\rm \Lsolar\,\Msolar^{-1}}$
\citep{scoville2001}.  \citet{andrews2011} expressed the expected $L_{\rm
IR}-L'_{\rm CO}$ and $L_{\rm IR}-L'_{\rm HCN}$ relations in the case of
Eddington-limited SFRs and found that, for CO luminosity tracing only the
actively star forming gas, the maximal possible luminosity is given by $L_{\rm
Edd}=4\pi G c \kappa^{-1} X_{\rm CO} L'_{\rm CO}$, where $\kappa$ is the
Rosseland-mean opacity, and $X_{\rm CO}$ is the $L'_{\rm CO}$-to-$M_{\rm H_2}$
conversion factor. A similar expression holds for HCN, albeit with different
$\kappa$ and $X$ values (see \citet{andrews2011} for details). Although the
exact normalization of this relation for each molecular line depends on poorly
constrained quantities like $\kappa$ and $X$, the Eddington limit set by the
strong FUV/optical radiation from embedded SF sites acting on the accreted dust
and dense gas can naturally provide the normalization of the observed $L_{\rm
FIR}-L'_{\rm CO}$ and $L_{\rm FIR}-L'_{\rm HCN}$ relations. In fact, adopting
$\kappa=5-30\,{\rm cm^2\,g^{-1}}$ and $X_{\rm CO}=0.8\,{\rm
K\,km\,s^{-1}\,pc^2}$, which are perfectly reasonable values for (U)LIRGs
\citep{thompson2005,solomon1997}, we find $\beta = \log(4\pi G c
\kappa^{-1}X_{\rm CO}) = 2.5-3.3$. We note that the high $\kappa$-value
($30\,{\rm cm^2\,g^{-1}}$), which corresponds to a three-fold increase in the
dust-to-gas mass ratio for the Rosseland-mean opacity (see \citet{andrews2011}
for details), as might be expected in (U)LIRGs, yields $\beta=2.5$, which is close
to the observed normalisation values obtained ($\beta \simeq 2$)
for the low-$J$ CO lines in \S~\ref{section:analysis} (Table \ref{table:fits};
see also Fig.\ \ref{figure:fig3}).

Given that the Eddington limit is ultimately set within individual SF sites
embedded deep inside molecular clouds, it will operate on all galaxies, not just
(U)LIRGs. For ordinary star forming spirals, the global $\beta$ normalization
value of the $L_{\rm FIR}-L'_{\rm CO}$ relations for low-$J$ CO lines will be
lower than its (Eddington limit)-set value by a factor approximately equal to
the logarithm of its dense gas fraction, i.e., $\log(f_{\rm dense})=\log(M_{\rm
dense}/M_{\rm tot})$. By the same token, the offset in the $\log( L_{\rm FIR}) -
\log( L'_{\rm CO})$ plane between two populations with significantly different
dense gas fractions ($f_{\rm dense, 1}$ and $f_{\rm dense, 2}$, say) can be
shown to be $\Delta \beta \sim \log( f_{\rm dense, 2}/f_{\rm dense, 1})$.  Thus,
an increasing $f_{\rm dense}(L_{\rm IR})$ function can cause the super-linear
FIR$-$CO(low-$J$) relations seen in some galaxy samples (which in reality is a
varying $\beta(L_{\rm IR})$ rather than a superlinear $\alpha$).

Local (U)LIRGs and high-$z$ DSFGs on the other hand form stars closer to the
Eddington limit on a global scale \citep{andrews2011}. Thus, the linear FIR$-$CO
(low-J) and FIR$-$HCN/CS relations observed for local (U)LIRGs and high-$z$
DSFGs is consistent with the notion that radiation pressure is an important physical
mechanism that underlies the observed star formation laws in highly
dust-obscured galaxies. In effect, the extreme merger/starbursts that dominate
the (U)LIRGs and high-$z$ DSFG population resemble dramatically scaled-up
versions of dense gas cores hosting SF deep inside Giant Molecular Clouds
(GMCs), with the balance between radiation pressure and self-gravity setting
their equilibrium during their IR-luminous phase.

In this framework the failure of the available theoretical models to account
for the observed $L_{\rm FIR}-L'_{\rm CO}$ relations of low-$J$ CO and heavy
rotor molecular lines might be attributed to the role radiation pressure
feedback plays in ultimately determining such relations.  This has not be taken
into account in all current theoretical considerations that either seek to
explain the S-K relation in galaxies (e.g., \citet{elmegreen2002}), or use an
(S-K) relation of $\rho_{\rm SFR}\propto (\rho_{\rm gas})^{1.5}$ to determine
the emergent $L_{\rm FIR}-L'_{\rm line}$ relations for molecular gas
\citep{krumholz2007,narayanan2008}. These use self-gravity, the associated
time-scale of free-fall time, along with models on how the SF efficiency (gas
mass fraction converted into stars per free fall time) varies per phase in
turbulent gas as the main ingredients towards a complete understanding of SF,
S-K relations, and the proxy $L_{\rm FIR}-L'_{\rm X}$ relations. A
non-gravitational force like that exerted by radiation pressure on accreted gas
and dust near SF sites can greatly modify such a picture by reducing or
eliminating the dependence on the free fall time, especially for the
high-density gas (which presumably is the one closest to active SF sites).
Alternatively, is has been suggested that in lower luminosity systems the star
formation may be regulated by feedback-driven turbulence (kinetic momentum
feedback) rather than by radiation pressure \citep{ostriker2011,shetty2012,kim2013}. 
Assuming a continuum optical depth at FIR wavelengths ($\tau_{\rm FIR}$) of
order unity for our sample of starburst/merger (U)LIRGs and typical dust
temperatures of $\sim 50\,{\rm K}$, we can make a rough estimate of the expected
radiation pressure, namely $P_{\rm rad} \sim \tau_{\rm FIR} \sigma T_{\rm d}^4/c
\sim 1.2\times 10^{-8}\,{\rm erg\,cm^{-3}}$ (where $\sigma$ is
Stefan-Boltzmann's constant and $c$ the speed of light). This is comparable to
the turbulent pressure $P_{\rm turb}\sim \rho \sigma_{\rm v}^2/3 \sim 1.4\times
10^{-8}\,{\rm erg\,cm^{-3}}$, obtained assuming a turbulent velocity dispersion
of $\sigma_{\rm v}\sim 5\,{\rm km\,s^{-1}}$ and an average gas mass density of
$\rho \sim 2\mu n_{\rm H_2}$ corresponding to $n_{\rm H_2}\sim 10^5{\rm
cm^{-3}}$.  Both of these greatly exceed the expected thermal pressure $P_{\rm
th}\sim n_{\rm H_{\rm 2}} k_{\rm B} T_{\rm k}\sim 1.4\times 10^{-9}\,{\rm
erg\,cm^{-3}}$ (for $n_{\rm H_2}\sim 10^5\,{\rm cm^{-3}}$ and $T_{\rm k}\sim
100\,{\rm K}$) -- thus highlighting the point made above that
a complete physically model of the S-K relations has to incorporate
the effects of radiation pressure and/or turbulence.

\section{The $\alpha$ and $\beta$ turnovers for high-$J$ CO lines}\label{section:alpha-and-beta}
Higher than $J=6-5$ neither the slope, $\alpha$, nor the normalization,
$\beta $, of the $L_{\rm FIR}-L'_{\rm CO}$ relations remain constant but
$\alpha$ decreases while $\beta$ increases towards higher $J$ levels.  This can
be understood using a simple argument first put forth (in a slightly different
form than here) by \citet{wong2002}. Consider that $\alpha_{{\rm CO}_{J,J-1}} =
d\log L_{\rm FIR}/d\log L'_{{\rm CO}_{J,J-1}}$ can be expressed as:
\begin{eqnarray}
\alpha_{{\rm CO}_{J,J-1}} &=& \frac{d\log L_{\rm FIR}}{d\log L'_{{\rm HCN}_{1,0}}}\times \frac{d\log L'_{{\rm HCN}_{1,0}}}{d\log L'_{{\rm CO}_{J,J-1}}}\\
       &=& \alpha_{{\rm HCN}_{1,0}} \left (1 + \frac{d\log l_{{\rm dense}_{J,J-1}}}{d\log L'_{{\rm CO}_{J,J-1}}}\right ),
\label{equation:slope}
\end{eqnarray}
where $\alpha_{{\rm HCN}_{1,0}}$ is the slope of the FIR$-$HCN $J=1-0$ relation,
which as mentioned previously, is near unity. The last term, $l_{{\rm
dense}_{J,J-1}} = L'_{{\rm HCN}_{1,0}}/L'_{{\rm CO}_{J,J-1}}$, is a convenient
parametrization of deviations in $\alpha_{{\rm CO}_{J,J-1}}$ from unity, and
depends on both the dense gas content (as traced by HCN) and the global CO line
excitation.

\begin{figure}[t]
\includegraphics[width=0.48 \textwidth]{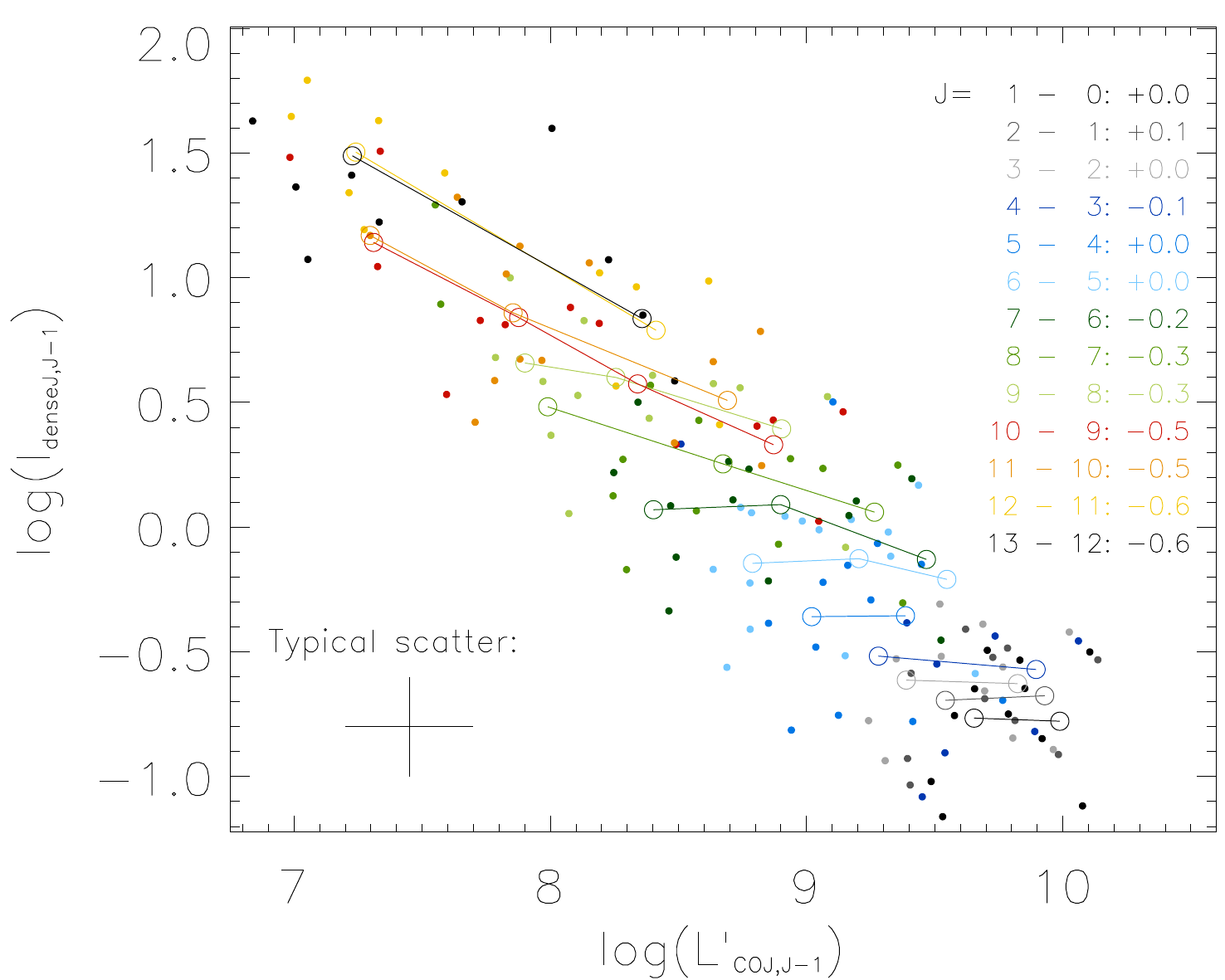}
\caption{$\log l_{{\rm dense}_{J,J-1}}$ vs.\ $\log L'_{{\rm CO}_{J,J-1}}$ for
our local (U)LIRG sample defined in \S~\ref{section:samples-and-data} (solid
symbols), where $l_{{\rm dense}_{J,J-1}} = L'_{\rm HCN_{1,0}}/L'_{\rm
CO_{J,J-1}}$ (see \S~\ref{section:alpha-and-beta}).  The different CO
transitions are color-coded (see insert). To highlight the trends, we show the
average $\log l_{{\rm dense}_{J,J-1}}$-values within suitable bins of $L'_{{\rm
CO}_{J,J-1}}$ (shown as open circles and connected with solid lines). Linear
fits to these averages yield the slopes given in the insert. For CO($1-0$) to
CO($6-5$), we find $d\log l_{{\rm dense}_{J,J-1}}/d\log L'_{{\rm CO}_{J,J-1}}
\simeq 0$, which when inserted in eq.\ \ref{equation:slope} yields $\alpha_{{\rm
CO}_{J,J-1}}\simeq 1$, in agreement with our findings. For higher CO
transitions, we have $d\log l_{{\rm dense}_{J,J-1}}/d\log L'_{{\rm CO}_{J,J-1}}
< 0$, which result in the sub-linear $\alpha_{{\rm CO}_{J,J-1}}$-values, which
match our directly determined FIR$-$CO slopes.}
\label{figure:fig4}
\end{figure}

For CO $J=1-0$, $l_{{\rm dense}_{1,0}}$ is a linear proxy of dense gas mass
fraction. This is simply due to the linearity of the $L'_{{\rm
HCN}_{1,0}}-L'_{{\rm CO}_{1,0}}$ relation (a fit to the $L_{\rm IR[8-1000\,{\rm \mu m}]}
> 10^{11}\,\Lsolar$ sources in the \citet{gao2004b} sample
yields $\log L'_{{\rm HCN}_{1,0}} \simeq 0.9\log L'_{{\rm CO}_{1,0}} - 0.2$), and
the fact that CO $J=1-0$ provides a good linear measure of $M_{\rm tot}({\rm
H}_2)$. The same applies also for the $J=2-1$ line. For the higher $J$ CO lines
$n_{\rm crit}$ becomes similar to that of HCN $J=1-0$ (or only slightly
surpasses it) while their $E_{J}/k_{\rm B}$ ($\sim 115-500\,{\rm K}$)
significantly exceed that of HCN $J=1-0$ ($\sim 4.3\,{\rm K}$). The
high-$J$ CO lines are significantly excited (see \S~\ref{subsection:CO-SLED})
and, following the argument first made by \citet{bradford2003}, this is unlikely
to be a pure density effect, as this would imply too large CO $J=1-0$ and $2-1$
optical depths and, in turn, $^{12}$CO$/^{13}$CO line ratios well below the
typical values ($\sim 10-30$) observed for local (U)LIRGs (e.g.,
\citet{casoli1992,aalto1995}). Instead, we argue that the high-$J$ CO lines are
produced by a dense and warm ($T_{\rm k} \gs 100\,{\rm K}$) phase. The $l_{{\rm
dense}_{J,J-1}}$ then becomes a measure of the $R_{\rm d,d-w}=M_{{\rm
dense}}({\rm H}_2)/M_{\rm dense-warm}({\rm H}_2)\geq 1$ modulo gas excitation
differences between the dense (d) and the dense and warm (d-w) molecular
gas reservoirs.  The derivative inside the parenthesis in eq.\
\ref{equation:slope} will be nearly zero for both low- and high-$J$ CO lines as
long as: a) the dense gas mass fraction remains nearly constant within our
galaxy sample (i.e., the sample is homogeneous in terms of $f_{\rm dense}$ and
its proxies $l_{{\rm dense}_{1,0}}$, $l_{{\rm dense}_{2,1}}$), and b) the
$R_{\rm d,d-w}$ ratio also remains constant. The latter means that the relative
excitation conditions and mass between the dense gas component (d) and its
sub-component of dense and warm gas (d-w) remain invariant across the sample.
The trend of $L_{\rm FIR}-L'_{\rm CO}$ relations above $J=6-5$ towards
increasing sub-linearity for higher $J$ levels is due to a decrease of $l_{{\rm
dense}_{J,J-1}}$ with increasing high-$J$ CO luminosity, thus resulting in
$\alpha_{{\rm CO}_{J,J-1}} < 1$. This behavior is indeed obvious in Fig.\
\ref{figure:fig4}, which shows $l_{{\rm dense}_{J,J-1}}$ as a function of
$L'_{{\rm CO}_{J,J-1}}$ for a sub-set of our local (U)LIRG sample with HCN($1-0$)
detections from \citet{gao2004a}. Note, we have not included DSFGs in this plot
since most detections of HCN at high redshifts are of QSOs and AGN dominated
DSFGs. In conjunction with eq.\ \ref{equation:slope}, Fig.\ \ref{figure:fig4}
can account for our established FIR$-$CO slopes in Fig.\ \ref{figure:fig2} and
Table \ref{table:fits}.

A decreasing $l_{{\rm dense}_{J,J-1}}$ with increasing high-$J$ CO luminosity
(yielding a negative derivative inside the parenthesis in eq.\
\ref{equation:slope}) indicates an increasing mass and/or excitation conditions
of the warm and dense (d-w) gas component relative to the dense gas reservoir
(d) that presumably contains it. This is possible if galaxies with increasingly
larger high-$J$ CO line luminosities (and thus also SFRs) increasingly have a
warm and dense gas component no longer tied to their SF via the average
FUV/optical radiation field.  Such examples have been found for individual
starbursts or star forming galactic nuclei
\citep{bradford2003,ward2003,hailey-dunsheath2008,
panuzzo2010,vanderwerf2010,rangwala2011,meijerink2013,rosenberg2014}, while the
presence of large masses of such a molecular gas component was recently
suggested as a general feature of the ISM in extreme merger/starbursts
\citep{papadopoulos2012}. High cosmic ray (CR) energy densities and/or the
dissipation of galaxy-wide shocks due to strong supersonic turbulence can
maintain $T_{\rm k}\ga 100\,{\rm K}$ for large amounts of high-density gas even
in the absence of FUV radiation fields (e.g., \citet{ao2013}).  Appreciable
fractions of dense gas mass per GMC above such temperatures demand different
heating mechanisms that can strongly heat the gas without readily dissociating
CO as FUV radiation does, and without being attenuated by dust (i.e., CR- and
turbulent heating). The onset of increasing normalization factors, $\beta$, of
the $L_{\rm FIR}-L'_{\rm CO}$ relations above $J=6-5$ is then simply another result of
the weakening link between the FUV-powered $L_{\rm FIR}$ and the thermal state of
dense gas for systems with high SFRs (and high-$J$ CO line luminosities). The
rapid rise of $\beta $ with $J$-level is expected if CO lines at increasingly
higher-$J$ levels probe ever higher gas thermal states with smaller mass per IR
luminosity.

The above picture retains the simple explanation for the observed $\alpha \simeq
1$ for the $L_{\rm FIR}-L'_{\rm CO}$ relations from $J=1-0$ to $J=5-4$, and for
those found for several heavy rotor molecular lines - as long as all these
relations refer to a near $f_{\rm dense}$-homogeneous galaxy sample, with SF
powering both the dust continuum and the molecular line luminosity via FUV
radiation. Highly super-linear slopes can only occur for galaxy samples with
significantly different dense gas fractions, or different star formation
relation normalizations (e.g., \citet{gao2004b}). Finally, in this overall
scheme, and for good (i.e., linear) dense SF gas tracers such as HCN and CS
lines it is rather hard to envisage how sub-linear slopes can come about
\citep{juneau2009}, since even the high-$J$ transitions of these heavy-rotor
molecules will trace the dense, cold star forming gas. Thus, the second term in
eq.\ \ref{equation:slope} will remain close to zero, leaving the FIR$-$HCN (or
FIR$-$CS) relation linear. In fact, linear slopes are observed for transitions
as high as CS $J=7-6$ \citep{zhang2014}.

\subsection{The CO SLEDs and the thermal state of high density gas}\label{subsection:CO-SLED}
\begin{figure*}[t]
\begin{center}
\includegraphics[width=0.5 \textwidth]{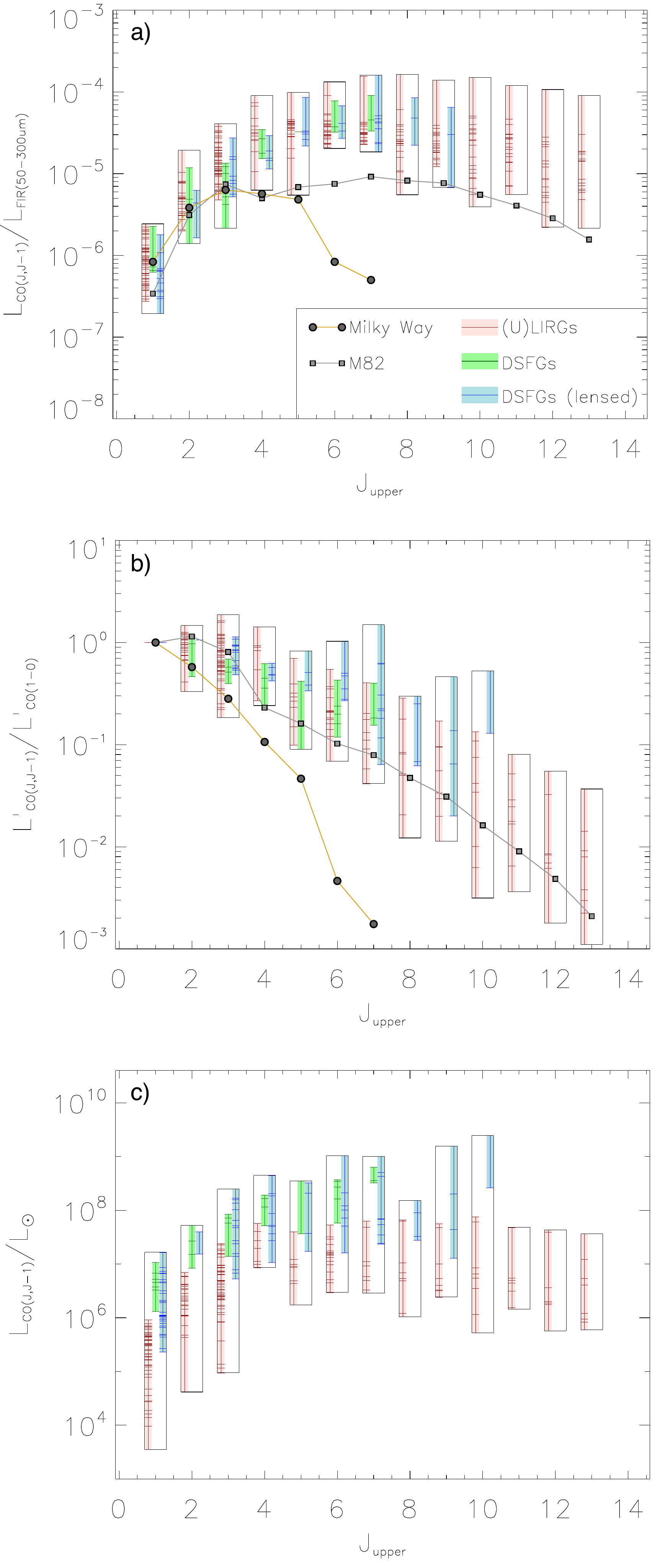}
\end{center}
\caption{The CO spectral energy distributions for the local (U)LIRG+HerCULES
sample (red), the unlensed (green) and strongly lensed (blue) high-$z$ DSFGs.
The CO SLEDs are given both as the CO line luminosities, in $\Lsolar$-units,
normalized by the FIR luminosity (top), as brightness temperature ratios, i.e.,
$L'_{\rm CO_{\rm J,J-1}}/L'_{\rm CO_{\rm 1,0}}$ (middle), and simply as $L_{\rm
CO_{\rm J,J-1}}$ (in $\Lsolar$-units) versus $J$ (bottom). The filled bars
indicate the full range of $L_{{\rm CO}_{J,J-1}}/L_{\rm FIR}$- and $L'_{\rm
CO_{\rm J,J-1}}/L'_{\rm CO_{\rm 1,0}}$-values in the two panels, respectively,
while the tick-marks indicate the values of individual sources. For comparison
we also show the CO SLEDs for the inner Galaxy (up to $J=7-6$) as measured  by
FIRAS/COBE \citep{fixsen1999}, normalised by $L_{\rm FIR}=(1.8\pm 0.6)\times
10^{10}\,{\rm \Lsolar}$ (also measured by FIRAS, \citet{wright1991}) and for
the proto-typical nearby starbust galaxy M\,82 \citep{panuzzo2010}.
}
\label{figure:fig5}
\end{figure*}

A more direct indication of significant amounts of warm and dense gas in our
(U)LIRG-dominated sample, and to what extent its thermal state is likely to be
maintained by the SFR-powered average FUV radiation fields, is provided by the
CO SLEDs. In Fig.\ \ref{figure:fig5} we show the FIR- and CO($1-0$)-normalised
CO SLEDs (top and middle panels, respectively), as well as the 'raw' CO
luminosities (bottom panel).  The first version allows for an assessment of the
CO SLEDs for the full samples (not all of our sources have CO $J=1-0$
measurements), and shows the cooling power of the CO lines with respect to the
continuum. The CO $J=1-0$ normalized representation of the CO SLEDs makes for a
direct comparison with observed CO line ratios in the literature, and is
furthermore what is usually used to constrain the excitation conditions of the
gas.

A detailed analysis of the CO SLEDs, in conjunction with the multi-$J$ HCN, CS
and HCO$^{+}$ line data-sets available for many of the (U)LIRGs in Fig.\
\ref{figure:fig5}, is needed for a full understanding of the heating and cooling
mechanisms of the molecular gas and for quantifying the relative mass-fractions
of the gas phases. Nevertheless, the marked contrast between their CO SLEDs and
that of the Milky Way disk (where most of bulk of the molecular gas is warmed by
photoelectric heating induced by the ambient FUV radiation field), already
indicates the presence of a different heating source.  While intense X-ray
radiation fields ($1-5\,{\rm keV}$) generated by AGN can penetrate and heat gas
up to $\gs 100\,{\rm K}$ at column densities of $10^{22}-10^{24}\,{\rm cm^{-2}}$
(and without dissociating all of the CO), it is unlikely to be the case here since
great care has been taken in removing AGN from our sample.  The integrated power
emitted in the CO $J=7-6$ to $J=13-12$ transitions for all the (U)LIRGs in our
sample, constitutes the bulk (about 60\,\%) of the total energy output of all
the CO lines. Exploring the effects of different CR and mechanical heating
rates on the thermal structure of clouds, \citet{meijerink2011} found that even
for extreme CR fluxes ($\sim 10^2-10^4\times$ the Milky Way value) it is
difficult to maintain temperatures $\gs 100\,{\rm K}$, and the effect on the
high-$J$ CO lines appears to be minor. Mechanical heating, such as supernova
driven turbulence and shocks, however, was found to heat the gas more
efficiently, and we favor this as the most likely explanation for hot
gas and the `boosted' high-$J$ CO lines observed in our (U)LIRGs.

Highly excited CO SLEDs have been found for the merger/starburst NGC\,6240, where a recent
analysis found FUV photons (and the resulting photoelectric heating) to be
inadequate as the main heating source for the high temperatures of its dense gas
(\citet{meijerink2013,papadopoulos2014}), and for Mrk\,231 where
X-rays from the AGN are thought to heat the dense molecular gas reservoir
\citep{vanderwerf2010} (although, Mrk\,231 has recently been shown to be much
less X-ray luminous than previously though, see \citet{teng2014}).  While such
CO SLEDs, and the need for alternative heating mechanisms than
FUV-photons to explain them, might be linked to the unusually high CO
line-to-continuum ratios of both of these two sources, a similar conclusion was
reached for M\,82 and NGC\,253 based on analyses of their full CO SLEDs
\citep{panuzzo2010,rosenberg2014}. Our work is the first to demonstrate such highly excited
high-$J$ CO SLEDs as a near generic characteristic of merger/starbursts (the
galaxies that dominate the sample shown in Fig.\ \ref{figure:fig5}).

\subsection{Some possible caveats}\label{subsection:caveats}
As mentioned in \S~\ref{section:samples-and-data} incorrect FIR$-$CO relations
may be inferred if the FIR and CO measurements cover different regions within
galaxies. This can be a serious problem for local extended sources where
single-dish CO beams can be smaller than the extent of the IR emission (see
discussion in \citet{zhang2014}).  We are confident that this is not an issue
for our HerCULES sample, where all SPIRE-FTS CO line fluxes were scaled to a
common $42\arcsecs$ angular resolution, which is beyond then the extent of the
IR emission in these sources (as traced by LABOCA $870\,{\rm \mu m}$ maps).
However, if this was not the case, and the SPIRE-FTS measurements did not
capture all the CO emission, it would imply that the derived FIR$-$CO slopes are
biased high (since the CO luminosity will be underestimated relative to the
total FIR luminosity, see Fig.\ \ref{figure:fig1}).  In short, the sub-linear
FIR$-$CO slopes at high-$J$ transition found here are robust against the
(unlikely) possibility that some (small) fraction of the CO emission is
unaccounted for.

The effects that the presence of strong AGNs would have on the SF relations are
two-sided. On the one hand it could lead to an overestimate of the IR luminosity
attributed to star formation, thus biasing the FIR$-$CO slopes high. On the
other hand, AGN-dominated environments, where penetrating X-rays may be
dominating the gas heating, tend to have `boosted' high-$J$ CO lines compared to
star forming regions (e.g., \citet{meijerink2007}).  If the AGN was deeply
buried it would not be easily detectable in X-rays and would be optically thick
in the IR, possibly down to mm wavelengths. The effect could be just what is
observed here -- a change of slope in the correlation and the high-$J$ CO lines
reflecting a hot, deeply embedded AGN. As mentioned in
\S~\ref{section:samples-and-data}, we used $L_{\rm FIR[50-300\,\mu m]}$ instead
of $L_{\rm IR[8-1000\,\mu m]}$ in the FIR$-$CO relations in order to minimize
the effects of AGN. More importantly, we only included local (U)LIRGs for which
the bolometric AGN contribution was deemed to be $<20\%$, as estimated from
several MIR diagnostics (\S~\ref{section:samples-and-data}). In the case of
the high-$z$ sample, obvious AGN-dominates systems were discarded from the
sample to begin with (\S~\ref{section:samples-and-data}).  Furthermore, we note
that deep X-ray observations as well as MIR spectroscopy of millimeter and
sub-millimeter selected DSFGs (which show no obvious signs of harboring an AGN)
have shown that any AGN that might be present typically contribute $\ls 20\,\%$
to the total IR luminosity \citep{alexander2005,mendez-delmestre2007}. Based on
the above, we feel confident that neither the FIR nor the CO luminosities are
biased high due to AGN, and that therefore our findings are not systematically
affected by AGN. 

Galaxies that are gravitationally lensed are prone to differential
magnification, an effect in which regions within a galaxy are magnified by
different amounts due to variations in their location within the galaxy, and/or
spatial extent \citep{blain1999}. This can significantly skew the observed
relative contributions from hot vs.\ cold dust to the IR luminosity, as well as
low- vs.\ high-$J$ CO line luminosity ratios \citep{serjeant2012}.  Furthermore,
a flux-limited sample of strongly lensed sources will tend to preferentially
select compact sources \citep{hezaveh2012}, which may be more likely to have
extreme CO excitation conditions. From Fig.\ \ref{figure:fig5}a, however, there
is nothing to suggest that the lensed DSFGs have markedly different $L'_{{\rm
CO}_{J,J-1}}/L_{\rm FIR}$ values than the non-lensed and local (U)LIRGs. In
Fig.\ \ref{figure:fig5}b, however, we do see a few lensed DSFGs which have
markedly higher $L'_{{\rm CO}_{J,J-1}}/L'_{{\rm CO_{1,0}}}$ ratios at high-$J$
than the other samples. This is exactly what we would expect to see if these
sources were differentially lensed, and the high-$J$ lines tracing more compact
regions than the $J=1-0$ line, or if the lensing preferentially selects compact
sources (which would tend to have more extreme excitation conditions).  The
strongest argument against our analysis being affected by differential
magnification effects is the fact that the lensed DSFGs make up only a minor
fraction of our total number of galaxies. To verify that this was indeed the
case we fitted the FIR$-$CO relations without the lensed DSFGs. This resulted in
slopes nearly identical to the ones given in Table \ref{table:fits}, and fully
consistent within the errors. Thus, we conclude that our findings are not
affected in any significant way by differential magnification effects.


\section{Summary}\label{section:summary}
Utilizing {\it Herschel}/SPIRE-FTS observations of a statistically significant
sample of 23 local (U)LIRGs, simultaneously covering the CO $J=5-4$ to
$J=13-12$ lines in one single spectrum, and combining these with CO $J=1-0$,
$2-1$, $3-2$ and $4-3$, $6-5$ data from our comprehensive ground-based CO survey
of the same sample, as well for an additional 44 local (U)LIRGs, we have
presented FIR$-$CO luminosity relations for the full CO rotational ladder from
$J=1-0$ to $J=13-12$. Included in our analysis is also a carefully groomed
sample of 35 high-$z$ lensed and unlensed DSFGs (spanning the redshift range $z\sim
1-6$) with robust FIR and CO luminosity measurements.  Due to their high
redshifts many of these sources have been observed in the mid- to high-$J$ CO
lines from the ground, thus allowing us to extend the mid- to high-$J$ FIR$-$CO
relations to the highest redshifts. 

For this data-set of low- and high-$z$ merger/starburst dominated galaxies we
find linear ($\alpha \simeq 1$) FIR$-$CO relations for CO $J=1-0$ up to $J=5-4$,
and with nearly constant normalization ($\beta \sim 2$). In light of the
linear star formation relations found for HCN and CS (e.g.,
\citet{gao2004b,zhang2014}, both of which are {\it bona fide} tracers of dense
star forming gas, we have shown that our results are to be expected provided the
dense gas mass fraction does not change significantly within the sample.  Our
findings are also qualitatively consistent with models in which the star
formation in (U)LIRGs is regulated on a global scale by radiation pressure as
these predict linear $L_{\rm FIR}-L_{\rm mol}$ slopes for any
molecule/transition that traces star forming gas in a 'homogeneous' sample
(i.e., constant normalization) \citep{andrews2011}.

For CO $J=6-5$ and up to $J=13-12$ we find increasingly sub-linear slopes and
higher normalization constants, which we argue is due to these lines effectively
being detached from the star formation as they trace gas that is dense ($\gs
10^4\,{\rm cm^{-3}}$) but also radically warmer ($\gs 100\,{\rm K}$) than what
is typical for star forming gas. This dense and warm ISM component is reflected
in the global CO SLEDs of the (U)LIRGs, and indeed of the high-$z$ DSFGs, which
remain highly excited from $J=6-5$ up to $J=13-12$. This suggests that star
formation powered by FUV radiation fields is unlikely to be
responsible for maintaining the gas temperature, but instead alternative
heating sources are required. Mechanical heating via shocks/turbulence 
seems to be the most plausible alternative given its effectiveness
(compared to CRs) at driving the temperatures in clouds to the required levels
($\sim 100\,{\rm K}$).

Finally, we note that our derived FIR$-$CO relations are sufficiently tight,
especially for the high-$J$ lines, that they can predict the expected CO line
brightness of high-$z$ DSFGs, which in turn might be useful for planned
observations with ALMA. \\

\bigskip

\bigskip

\acknowledgments The authors gratefully acknowledge financial support under the
"DeMoGas" project.  The project “DeMoGas” is implemented under the "ARISTEIA"
Action of the "Operational Programme Education and Lifelong Learning". The
project is co-funded by the European Social Fund (ESF) and National Resources.
TRG acknowledges support from an STFC Advanced Fellowship. TRG was also
supported by Chinese Academy of Sciences Fellowship for Young International
Scientists (grant no.  2012y1ja0006). ZYZ acknowledges support from the European
Research Council (ERC) in the form of Advanced Grant,{\sc cosmicism}.  We are
indebted to P.P.\ Papadopoulos for extensive discussions and comments on the
paper ({\it He ho'okele wa'a no ka la 'ino}).  Basic research in infrared
astronomy at the Naval Research Laboratory is funded by the Office of Naval
Research. JF also acknowledges support from the NHSC/JPL.  The research
presented here has made use of the NASA/IPAC Extragalactic Database (NED) which
is operated by the Jet Propulsion Laboratory, California Institute of
Technology, under contract with the National Aeronautics and Space
Administration.  Finally, we would like to thank the anonymous referee for a
useful and constructive referee report, which helped improve the paper.

\bibliographystyle{apj}

\clearpage

\end{document}